\documentclass[preprint,onecolumn,nofootinbib]{revtex4-1}
\pdfoutput=1

\usepackage[colorlinks=true,linkcolor=blue,urlcolor=blue,filecolor=black,citecolor=red,pdfstartview=FitV,pdftitle={},pdfsubject={},pdfkeywords={},pdfpagemode=None,bookmarksopen=true]{hyperref}
\usepackage{graphicx}
\usepackage{amsmath}
\usepackage{amsfonts}
\usepackage{amssymb,ulem}
\usepackage{subcaption} 

\usepackage{color}%
\usepackage{dcolumn}
\usepackage[utf8]{inputenc}
\newcommand{\f}{\begin{equation}}
	\newcommand{\ff}{\end{equation}}
\newcommand{\fa}{\begin{eqnarray}}
	\newcommand{\ffa}{\end{eqnarray}}

\begin{document}
	\title{Chaos dynamics of charged particles near Gibbons-Maeda-Garfinkle-Horowitz-Strominger black holes}
	\author{Zhen-Meng Xu$^{1}$}
	\author{Da-Zhu Ma$^{2}$}
	\thanks{mdzhbmy@126.com, Corresponding author}
	\author{Kai Li$^{1}$}
	\affiliation{\small{$^1$~School of Mathematics and Statistics, Hubei Minzu University, Enshi 445000, China}}
	\affiliation{\small{$^2$~College of Intelligent Systems Science and Engineering, Hubei Minzu University, Enshi 445000, China}}

	\begin{abstract}

The Gibbons-Maeda-Garfinkle-Horowitz-Strominger (GMGHS) dilatonic black hole, a key solution in low-energy string theory, exhibits previously unexplored chaotic dynamics for charged test particles under electromagnetic influence. While characterizing such chaos necessitates high-precision numerical solutions, our prior research confirms the explicit symplectic algorithm as the optimal numerical integration tool for strongly curved gravitational celestial systems. Leveraging the Hamiltonian formulation of the GMGHS black hole, we develop an optimized fourth-order symplectic algorithm $PR{K_6}4$. This algorithm enables a systematic investigation of the chaotic motion employing four distinct chaos indicators: Shannon entropy, Poincaré sections, the maximum Lyapunov exponents, and the Fast Lyapunov indicators. Our results demonstrate a critical dependence of chaos on both electric charge ($Q$, characterized by the Coulomb parameter $Q^*$) and magnetic charge ($Q_m$). Specifically, in electrically charged backgrounds, order-to-chaos transitions arise with increasing $Q$ or decreasing $Q^*$. Conversely, in magnetically charged backgrounds, chaos emerges as $Q_m$ increases. These findings validate Shannon entropy as a robust chaos indicator within relativistic frameworks and provide novel insights on the dynamics of string-theoretic black holes.

\end{abstract}
	\maketitle
\section{Introduction}

String theory stands as the only currently viable self-consistent theoretical framework with the potential to quantize gravity and unify it with electromagnetic, weak, and strong interactions. It holds immense promise not only for explaining the origin and fundamental workings of the universe and addressing numerous modern physics conundrums, but its low-energy limit predictions also provide a unique platform for exploring quantum gravitational effects. Within this framework, Einstein-Maxwell-Dilaton-Axion (EMDA) theory serves as the low-energy effective description of heterotic string theory in four-dimensional spacetime. By incorporating couplings between the dilaton, axion, and electromagnetic fields, it reveals symmetries inherent in string theory and the quantum properties of black holes. Related research encompasses black hole accretion processes \cite{Banerjee:2020ubc,Feng:2023iha,Feng:2024iqj}, the Kerr-Sen black hole as a spin particle accelerator \cite{An:2017hby}, exact solutions of the Klein-Gordon equation in Kerr-Sen spacetime \cite{Wu:2003,Vieira:2018hij}, and the connection between quasinormal modes (QNMs) and the shadows of rotating Kerr-Sen black holes \cite{Wu:2021pgf,Cardoso:2008bp}. Therefore, it is of great importance to investigate EMDA theory and its predicted astrophysical phenomena, testing the theory as an alternative to general relativity \cite{Narang:2020bgo}.

Dilaton black holes arising from string theory exhibit significant differences from the predictions of general relativity due to the dilaton parameter. This distinctive behavior has attracted considerable research interest. Gibbons and Maeda \cite{GM} were the first to derive charged black hole solutions within Einstein-Maxwell-dilaton models. Subsequently, Garfinkle, Horowitz, and Strominger \cite{Garfinkle:1990qj} rederived these solutions in the context of string theory, designating them as the GMGHS solutions. Within the EMDA theoretical framework, Sen \cite{Sen:1992ua} discovered the Kerr-Sen solution—a rotating generalization of the GMGHS solution. This solution is characterized by three parameters: mass $M$, charge $Q$, and angular momentum per unit mass $a$. When the parameter $a$ vanishes, the Kerr-Sen solution reduces to the static, spherically symmetric GMGHS solution, which possesses a dilaton parameter given by $r_{0} = Q^{2}/M$. Furthermore, if the charge $Q$ also vanishes, the solution reduces to the vacuum Kerr solution.

The null and timelike geodesics for uncharged particles around GMGHS black holes have been extensively studied. For example, Soroushfar et al. \cite{Soroushfar:2016yea} presented analytical solutions and applications of geodesic equations in static and rotating dilaton black holes. Blaga \cite{Blaga:2014spa} investigated timelike geodesics around a spherically symmetric electrically charged dilaton black hole, classifying orbital types using the effective potential of free particles. Fernando \cite{Fernando:2011ki} employed heterotic string theory to examine the null geodesics of static charged black holes, providing a detailed analysis of geodesics in both Einstein and string coordinate systems. The results demonstrate that in Einstein coordinates, geodesics can be solved exactly using Jacobi elliptic integrals across all possible photon energy levels and angular momenta. Recently, Lungu et al. \cite{Lungu:2024ewz} further studied charged test particle motion around GMGHS dilaton black holes using effective potential analysis and Lagrangian dynamics. Their research revealed that stable bound orbits form around electrically charged black holes when the dilaton parameter satisfies $r_0 \approx 0.36M$. For magnetized black holes, particle motion is constrained to a Poincaré cone, with circular orbits existing depending on $r_0$. These results align closely with astrophysical observations constraining $r_0$ ($0.1M \lesssim r_0 \lesssim 0.4M$), providing key theoretical support for the classical kinematics of dilaton black holes.

In fact, the study of chaos in black holes represents a critical frontier in theoretical physics, bridging general relativity, quantum gravity, thermodynamics, and information theory. Its significance is multifaceted: as extreme laboratories where quantum effects and strong gravity coexist, chaotic dynamics reveal limitations of classical relativity and provide benchmarks for quantum gravity theories \cite{MA2022epjc,MA2020JHEP}. Chaos underpins resolutions to the information paradox by suggesting horizons scramble information into Hawking radiation, evidenced by OTOCs saturating the Maldacena-Shenker-Stanford bound \cite{Mss2016}. In holography (AdS/CFT), chaos tests how emergent spacetime relates to quantum entanglement, while thermodynamically, it connects via Kolmogorov-Sinai entropy to microstructure insights near phase transitions. Besides, chaos drives gravitational wave bursts from binary orbits, explains QPOs in accretion disks, tests cosmic censorship via instabilities in exotic solutions, and reveals universal behavior intrinsic to quantum gravity \cite{MA2014PRD}. Crucially, chaos generates distinct astrophysical imprints—notably in black hole shadows, where unstable photon orbits governed by chaotic dynamics produce observable fractal structures. This shadow chaos directly probes spacetime geometry and serves as a unique observational signature. Thus, chaos in black holes is far more than a dynamical curiosity. 

Based on this, we mainly focuses on the chaotic dynamics evolution of charged particles near GMGHS black holes in this paper. However, it has been reported that the chaotic dynamics evolution of particles heavily relies on the stability and accuracy of numerical solutions \cite{MA2022epjc, MA2008216}. Due to the multiphysics coupling and complex dynamical behavior inherent in GMGHS black holes, traditional numerical methods tend to accumulate energy errors during long-term integration, leading to spurious chaos \cite{MA2008216}. Thus, it is necessary to use structure-preserving algorithms to ensure computational reliability. Symplectic algorithms preserve the symplectic structure in phase space and rigorously conserve the Hamiltonian in long-term orbital integration. This overcomes the numerical distortion problem of traditional methods in extremely curved spacetime, making symplectic methods the preferred approach for Hamiltonian systems. Wang et al. \cite{Wang:2021gja,Wang:2021xww,Wang:2021yqk} demonstrated the symplectic method's superiority by successfully applying symplectic integrators to diverse curved spacetimes, including Schwarzschild, Reissner-Nordström (RN), and Reissner-Nordström-(anti)-de Sitter black holes. After obtaining reliable particle trajectories using symplectic algorithms, the key step is identifying and analyzing chaotic phenomena within them. Traditional chaos identification tools include Poincaré sections, Lyapunov exponents, and power spectra \cite{Gerald:1988,Li:2018wtz,Zhang:2023lrt}. Recently, Cao et al. \cite{Cao:2024bjk} proposed Shannon entropy as a novel criterion for chaotic particle motion in curved spacetime and validated its effectiveness in magnetized black hole spacetimes, providing a new approach for our research.

In this paper, we establish a chaotic dynamics analysis framework based on the $PRK_{6}4$ symplectic algorithm for charged and magnetized GMGHS black holes. We then use Shannon entropy as an indicator to study how charge and magnetic charge parameters modulate the chaos threshold through the dilaton parameter $r_0$, leading to chaotic behavior. Additionally, the results are further validated by other chaos indicators. The paper is structured as follows: Section \ref{section2} presents the GMGHS black hole geometry; Section \ref{section3} analyzes chaotic dynamics in electrically charged black holes; Section \ref{section4} discusses chaotic features in magnetically charged black holes; and Section \ref{section5} provides concluding remarks.

\section{The GMGHS Black Hole Solution}\label{section2}

The GMGHS solution is a classical solution under the low-energy approximation of string theory, providing an analytical model for studying string-theoretic black holes. In this limit, gravity couples to the dilaton and electromagnetic fields, enabling investigations into black hole thermodynamics and quantum gravitational effects.

In Boyer-Lindquist coordinates $(t,r,\theta ,\varphi)$, the static spherically symmetric charged dilatonic black hole line element is defined as
\begin{equation}\label{1}
	ds^2=-f(r)dt^2+f(r)^{-1}dr^2+p^2\left(d\theta^2+\sin^2\theta d\varphi^2\right).
\end{equation}
Where,
\begin{equation}\label{2}
	f(r)=1-\frac{2M}{r}, p^2=r(r-r_0), r_0=\frac{Q^2e^{2\phi_{0}}}{M}.
\end{equation}
Here, $M$ and $Q$ are the mass and charge of the black hole. The dilaton field $\phi$ satisfies $e^{-2\phi}=e^{-2\phi_{0}}-\frac{Q^{2}}{Mr}$, with $\phi_{0}$ being the asymptotic constant value of the dilaton. For simplicity, we set $\phi_{0}=0$, reducing the GMGHS metric to 
\begin{eqnarray}\label{3}
	ds^{2}&=&-\left(1-\frac{2M}{r}\right)dt^{2}+\left(1-\frac{2M}{r}\right)^{-1}dr^{2}\nonumber\\
	&&+r\left(r-\frac{Q^{2}}{M}\right)d\theta^2+r\left(r-\frac{Q^{2}}{M}\right)\sin^2\theta d\varphi^2.
\end{eqnarray}
Since the metric reduces to the Schwarzschild form at fixed $\theta$ and $\varphi$, this charged black hole possesses a regular event horizon at $r_h = 2M$. The parameter $r_0$ is constrained to $0 < r_0 < 2M$, where the extremal limit $r_0 = 2M$ corresponding to $Q^2 = 2M^2$. Throughout this work, we use geometrized units with $c=G=1$, and normalize both the black hole mass $M$ and test particle mass $m$ to unity ($m=M=1$). 

The vector potential for a magnetically charged black hole is $A_{\mu}=(0,0,0,-Q_{m}\cos\theta)$, while for an electrically charged black hole it is $A_{\mu}=(-Q/r,0,0,0)$. In subsequent analyses, we separately study the chaotic dynamics of charged test particles in external electric or magnetic fields around the GMGHS black hole.

\section{The electrically charged black hole} \label{section3}
\subsection{Hamiltonian system}
The motion of a test particle with charge $q$ orbiting a black hole is described by a Hamiltonian system. For a charged black hole with charge $Q$, where the electric field originates from the vector potential component $A_t=-Q/r$, the Lagrangian is
\begin{eqnarray}\label{4}
	\mathcal{L}&=&\frac{1}{2}g_{\mu\nu}\dot{x}^{\mu}\dot{x}^{\nu}+qA_{\mu}\dot{x}^{\mu} \nonumber\\
	&=&\frac{1}{2}\left[-f\dot{t}^2+\frac{\dot{r}^2}{f}+p^2(\dot{\theta}^2+\sin^2\theta\dot{\varphi}^2)\right]-\frac{Q^*\dot{t}}{r}.
\end{eqnarray}
Here, the dot denotes differentiation with respect to proper time $\tau$, and $f$ represents $f(r)$. Additionally, $Q^*= qQ$ parameterizes the electromagnetic Coulomb interaction. For a timelike particle, the four-velocity satisfies the normalization constraint:
\begin{equation}\label{5}
	g_{\mu\nu}\dot{x}^{\mu}\dot{x}^{\nu}=-1.
\end{equation}
The covariant generalized momentum is defined as
\begin{equation}\label{6}
	p_\mu=\partial\mathcal{L}/\partial\dot{x}^\mu=g_{\mu\nu}\dot{x}^\nu+qA_{\mu}.
\end{equation}
Due to the cyclic coordinates $t$ and $\varphi$, the energy $E$ and angular momentum $L$ are conserved:
\begin{eqnarray}\label{7}
	-{{E}} &=& {p_t} = {g_{tt}}\dot t + q{A_t}=-f\dot{t}-\frac{Q^*}{r},\nonumber\\
	{{L}} &=& {p_\varphi } = {g_{\varphi \varphi }}\dot \varphi  + q{A_\varphi }=p^2\sin^2\theta\dot{\varphi}. 
\end{eqnarray}
The Hamiltonian describing the particle's motion is thus
\begin{eqnarray}\label{8}
	H&=&p_{\mu}\dot{x}^{\mu}-\mathcal{L}=\frac{1}{2}g^{\mu\nu}(p_{\mu}-qA_{\mu})(p_{\nu}-qA_{\nu}) \nonumber\\
	&=&\frac{1}{2}\left[\frac{(Q^*-Er)^{2}}{r(2-r)}+\frac{L^{2}\csc^{2}\theta}{-Q^{2}r+r^{2}}\right]+\frac{(r-2)p_r^{2}}{2r}\nonumber\\
	&&+\frac{p_\theta^{2}}{2r(-Q^{2}+r)}. 
\end{eqnarray}
The Hamiltonian in Eq. \eqref{8} is constrained by the rest mass condition: 
\begin{equation}\label{9}
	H=-\frac{1}{2}.
\end{equation}

\subsection{Construction of Explicit Symplectic Integrators}

In the presence of an asymptotically uniform electric field, the Hamiltonian in Eq. \eqref{8} lacks a fourth motion integral, rendering the system non-integrable and potentially chaotic. Reliable numerical algorithms are therefore essential for studying particle chaotic dynamics. Compared to traditional methods (e.g., Runge-Kutta), symplectic integrators exhibit superior performance in long-term Hamiltonian integration. They preserve the symplectic structure, rigorously conserve the Hamiltonian, and eliminate spurious chaos induced by energy drift.

Adopting the decomposition scheme from \cite{Wang:2021gja,Wang:2021xww,Wang:2021yqk}, we partition the Hamiltonian \eqref{8} into four integrable components:
\begin{equation}\label{10}
	H = {H_1} + {H_2} + {H_3} + {H_4}.
\end{equation}
Where
\begin{eqnarray}\label{11}
	{H_1} &=& \frac{1}{2}\left[\frac{(Q^*-Er)^{2}}{r(2-r)}+\frac{L^{2}\csc^{2}\theta}{-Q^{2}r+r^{2}}\right],\nonumber \\
	{H_2} &=& \frac{p_r^2}{{2}},\nonumber \\
	{H_3} &=& \frac{{-p_r^2}}{r},\nonumber \\
	{H_4} &=& \frac{p_\theta^{2}}{2r(-Q^{2}+r)}.
\end{eqnarray}
Each sub-Hamiltonian admits an analytical solution (details in Appendix \ref{appendix}). Their solvers are denoted ${{{\cal H}}_1}$, ${{{\cal H}}_2}$, ${{{\cal H}}_3}$, and ${{{\cal H}}_4}$. Previous work shows the $PR{K_6}4$ algorithm achieves significantly higher accuracy than second-order ($S2$)and fourth-order Yoshida ($S4$) symplectic methods \cite{Xu:2024ble}. We therefore directly implement the $PR{K_6}4$ symplectic algorithm.

Letting $h$ be the time step, we define two first-order symplectic operators:
\begin{eqnarray}\label{12}
	{\chi _h} = {{{\cal H}}_4}(h) \times {{{\cal H}}_3}(h) \times {{{\cal H}}_2}(h) \times {{{\cal H}}_1}(h), 
\end{eqnarray}
\begin{eqnarray}\label{13}
	\chi _h^ *  = {{{\cal H}}_1}(h) \times {{{\cal H}}_2}(h) \times {{{\cal H}}_3}(h) \times {{{\cal H}}_4}(h).
\end{eqnarray}
where $\times$ denotes composition. The fourth-order optimal explicit $PRK$ symplectic algorithm($PR{K_6}4$) \cite{Zhou_2022} with first-order operators ${\chi _h}$ and $\chi _h^*$ is then
\begin{eqnarray}\label{14}
	PRK_64&=&\chi_{c_{12}h}\times\chi_{c_{11}h}^*\times\chi_{c_{10}h}\times\chi_{c_9h}^*\times\chi_{c_8h}\times\chi_{c_7h}^*\nonumber\\&&\times\chi_{c_6h}\times\chi_{c_5h}^*\times\chi_{c_4h}\times\chi_{c_3h}^*\times\chi_{c_2h}\times\chi_{c_1h}^*.
\end{eqnarray}
The time coefficients \cite{Zhou_2022} are 
\begin{eqnarray}
	\nonumber
	&&c_1=c_{12}= 0.079203696431196,   \\ \nonumber
	&&c _2=c_{11}= 0.130311410182166,   \\ \nonumber
	&&c_3=c_{10}= 0.222861495867608,    \\ \nonumber
	&&c_4=c_9=-0.366713269047426,    \\ \nonumber
	&&c_5=c_8= 0.324648188689706,   \\ \nonumber
	&&c_6=c_7=0.109688477876750.   \nonumber
\end{eqnarray}

\subsection{Chaos indicators}

Chaos describes the unpredictable, complex behavior that arises in deterministic systems due to extreme sensitivity to initial condition. To quantify and detect chaos, many chaos indicators are employed, such as Poincaré sections, the maximum Lyapunov exponent (MLE), the Fast Lyapunov indicators (FLI), the relative finite-time Lyapunov indicators (RLI), the smaller alignment indices (SALI), and the generalized alignment indices (GALI) \cite{MA2020JHEP}. Each technique possesses distinct strengths and limitations, Poincaré sections are restricted to four-dimensional phase spaces. MLE, FLI, RLI, SALI, and GALI primarily serve conservative systems like n-body problems, cosmology, and general relativity models, though only MLE functions effectively in dissipative contexts. MLE quantifies chaos by measuring trajectory divergence, calculable via variational or two-particle methods. Crucially, only coordinate-invariant MLE applies to Mixmaster cosmology and conservative Post-Newtonian spinning binaries. RLI operates faster than MLE and suits symplectic mappings or continuous Hamiltonians; both require a single deviation vector. SALI necessitates two vectors, while GALI—an enhanced SALI variant—delivers faster, more detailed local dynamics insights. FLI offers the greatest computational efficiency by avoiding limit calculations and variational equations. In fact, RLI, SALI, and GALI are variants of MLI, and their essence is the same. Recently, a new chaos indicator named Shannon entropy is proposed \cite{Cao:2024bjk}. 

Shannon entropy, a fundamental concept in information theory, serves as a measure for non-equilibrium systems:
\begin{equation}\label{15}
	H(X)=-\sum_{i=1}^{n}p(x_{i})\log_{b}p(x_{i}).
\end{equation}
Where $H (X)$ is the entropy, $p(x_{i})$ is the probability of event $x_{i}$, $n$ is the total number of events, and $b=2$.
In order to test the ability of Shannon entropy to detect chaos in strongly gravitational celestial systems, we use this method as the basis chaos indicators and compared its results with the results of Poincaré sections, MLE and FLI.

\subsection{Impact of Parameter Variations}

\begin{figure*}[htbp]
	\centering
	\captionsetup[subfigure]{labelformat=empty} 
	
	\begin{subfigure}[t]{0.32\textwidth}
		\includegraphics[width=\linewidth]{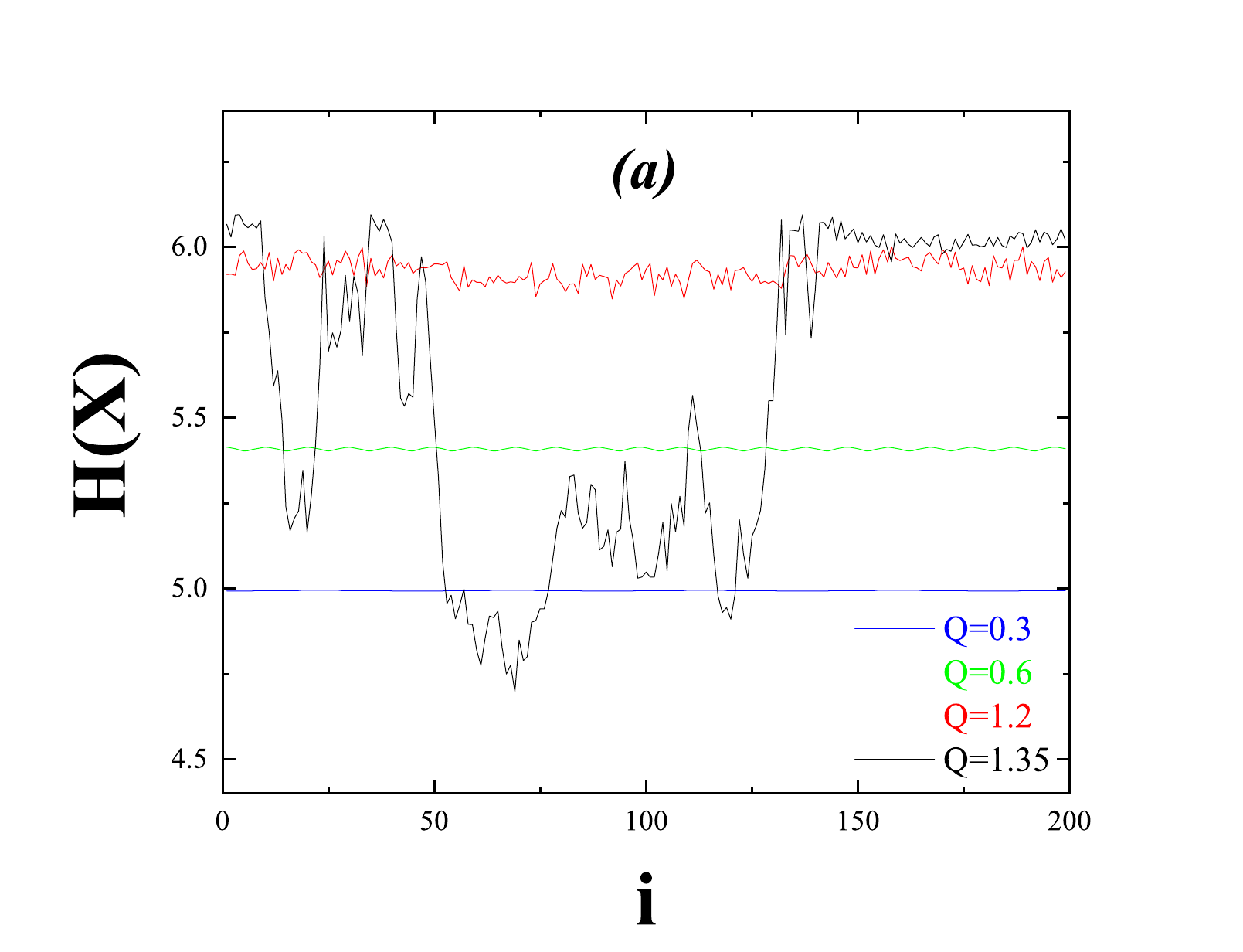}
		\phantomcaption 
		\label{fig:subfig1a}
	\end{subfigure}
	\hfill
	\begin{subfigure}[t]{0.32\textwidth}
		\includegraphics[width=\linewidth]{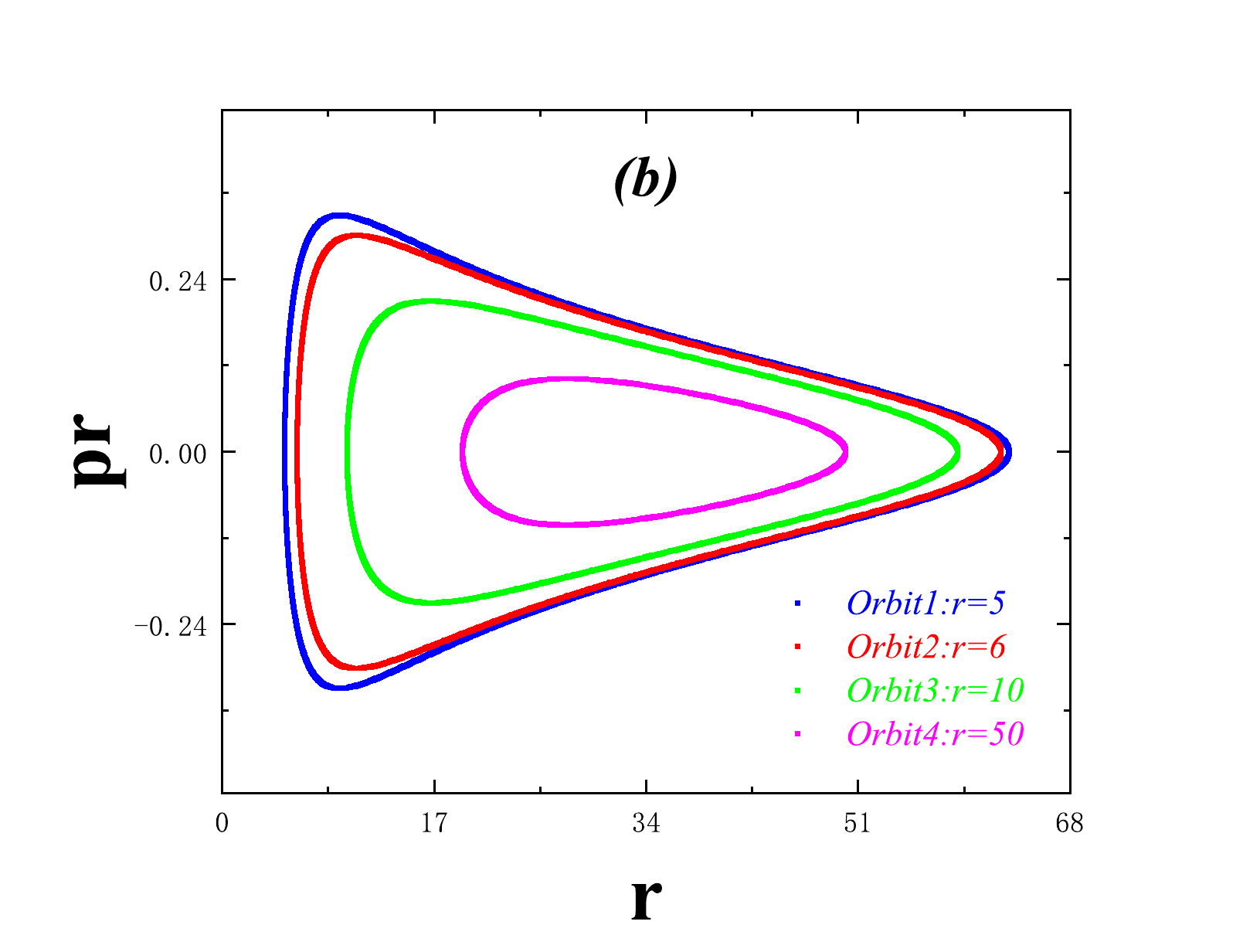}
		\phantomcaption
		\label{fig:subfig1b}
	\end{subfigure}
	\hfill
	\begin{subfigure}[t]{0.32\textwidth}
		\includegraphics[width=\linewidth]{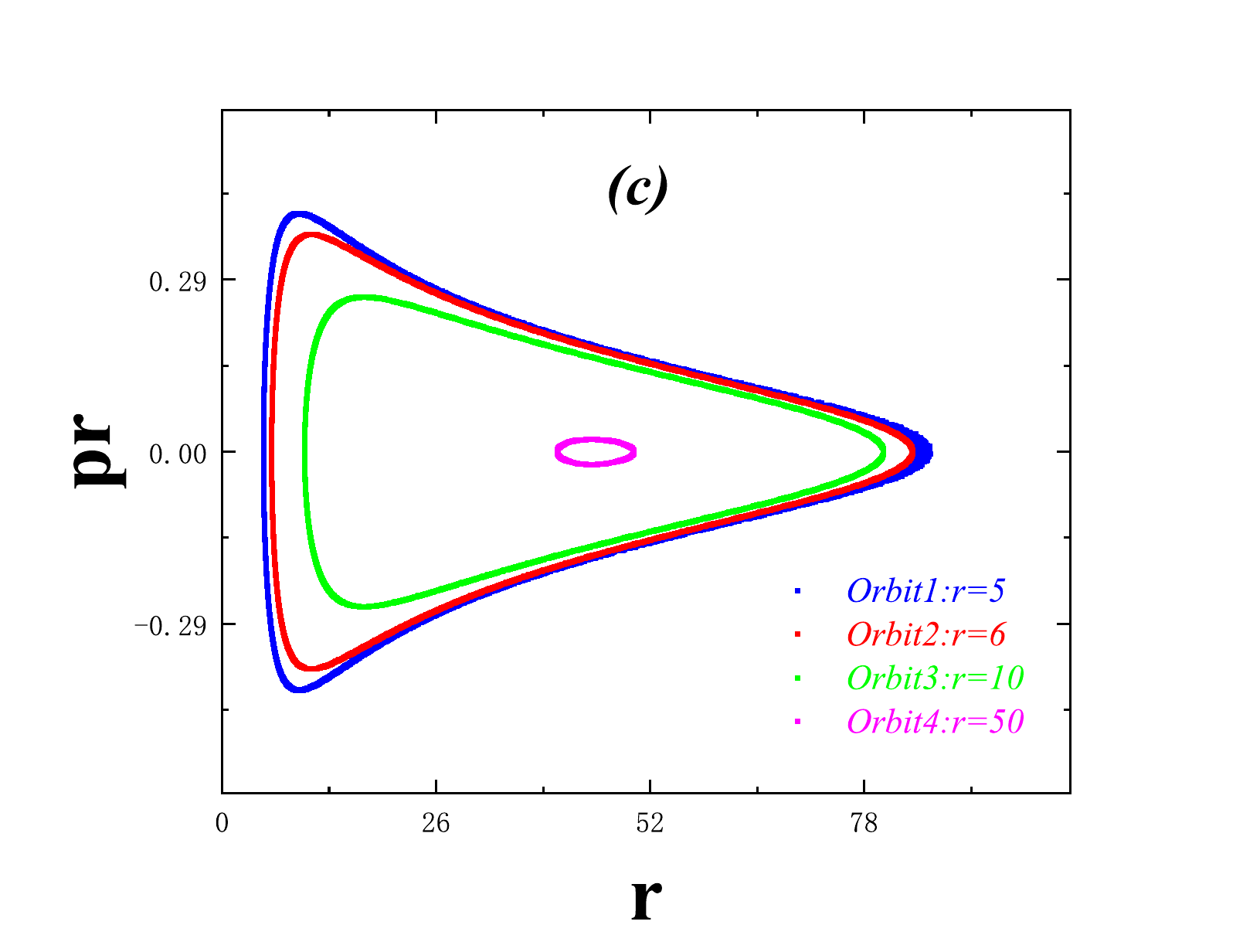}
		\phantomcaption
		\label{fig:subfig1c}
	\end{subfigure}
	
	\vspace{10pt} 
	\begin{subfigure}[t]{0.32\textwidth}
		\includegraphics[width=\linewidth]{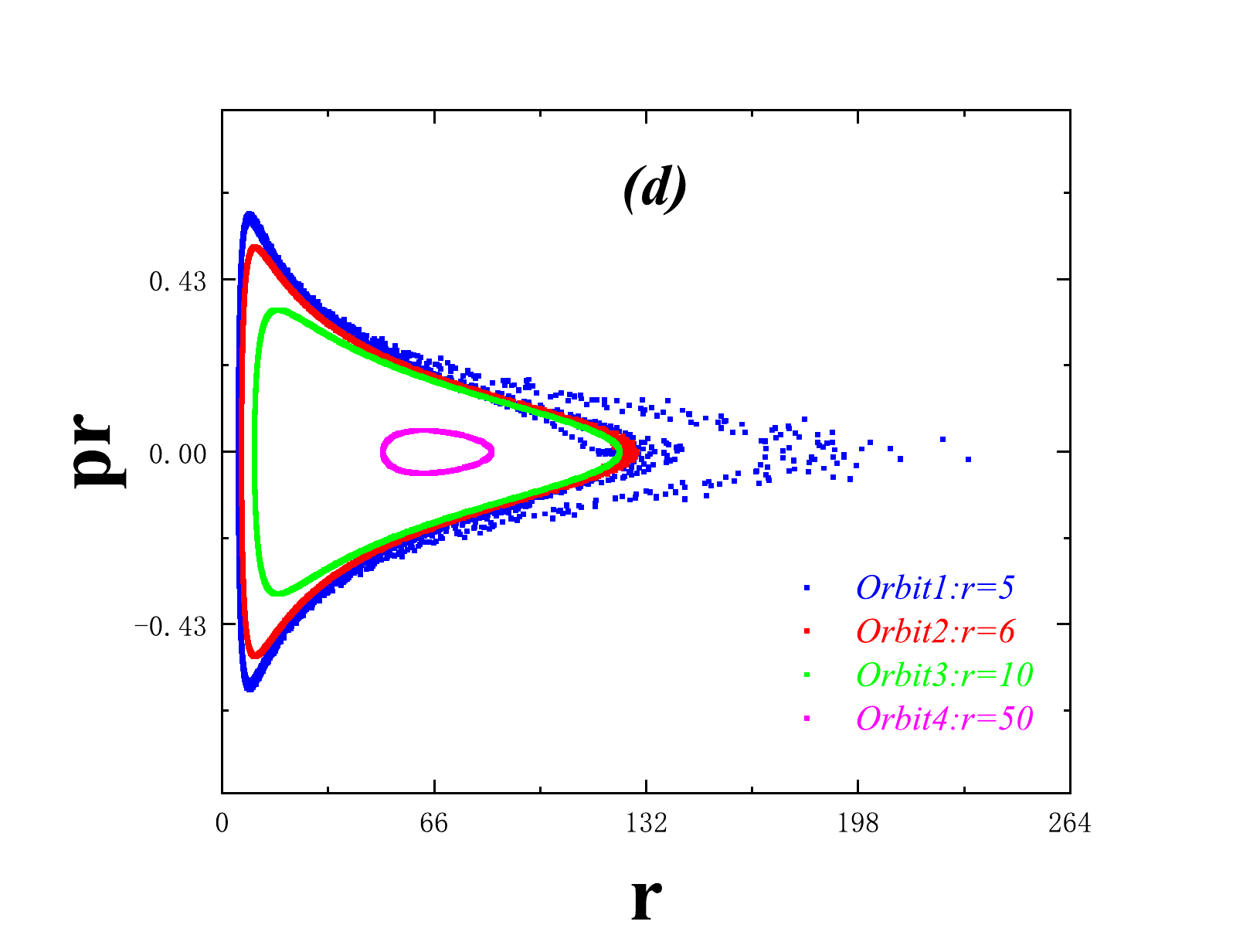}
		\phantomcaption
		\label{fig:subfig1d}
	\end{subfigure}
	\hfill
	\begin{subfigure}[t]{0.32\textwidth}
		\includegraphics[width=\linewidth]{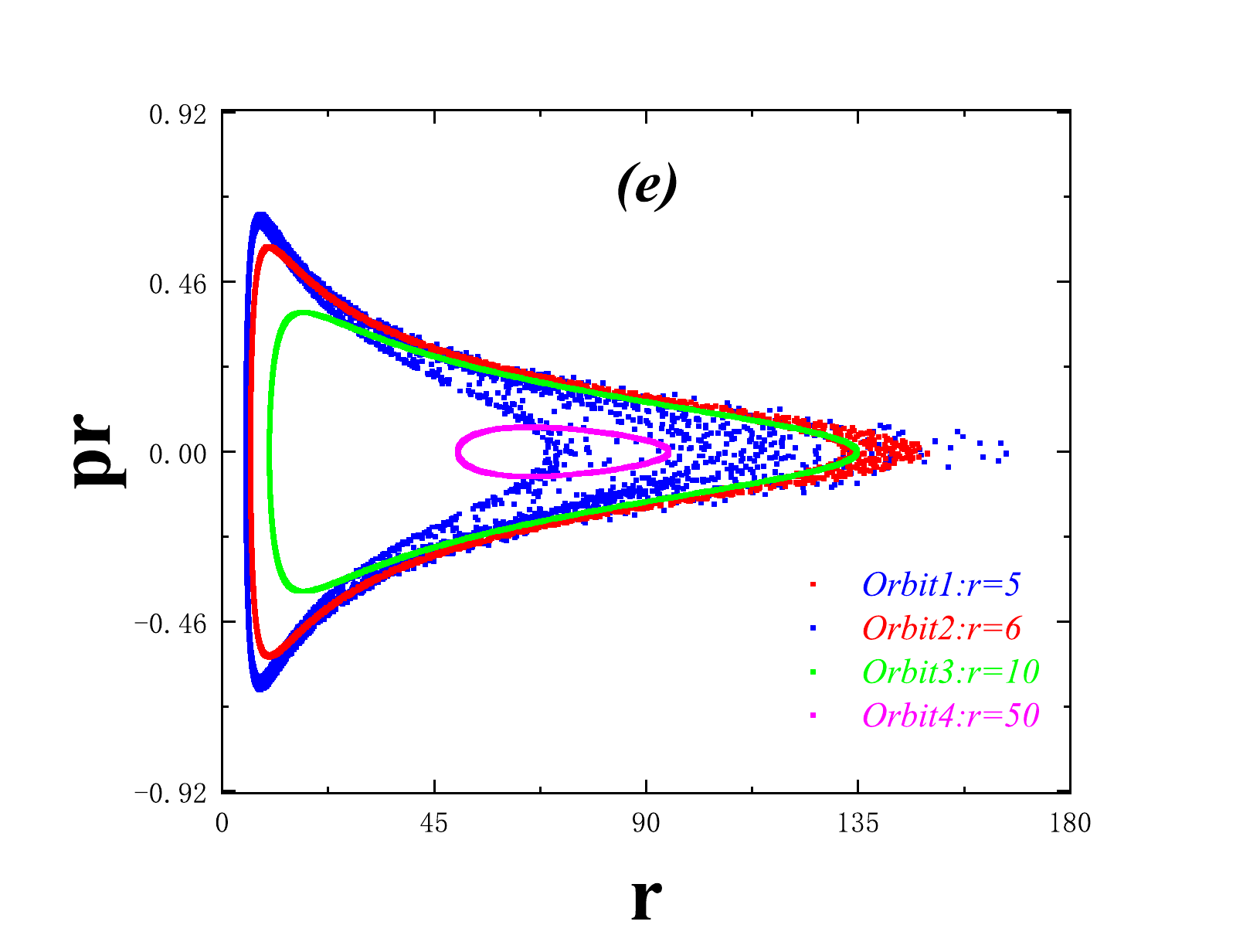}
		\phantomcaption
		\label{fig:subfig1e}
	\end{subfigure}
	\hfill
	\begin{subfigure}[t]{0.32\textwidth}
		\includegraphics[width=\linewidth]{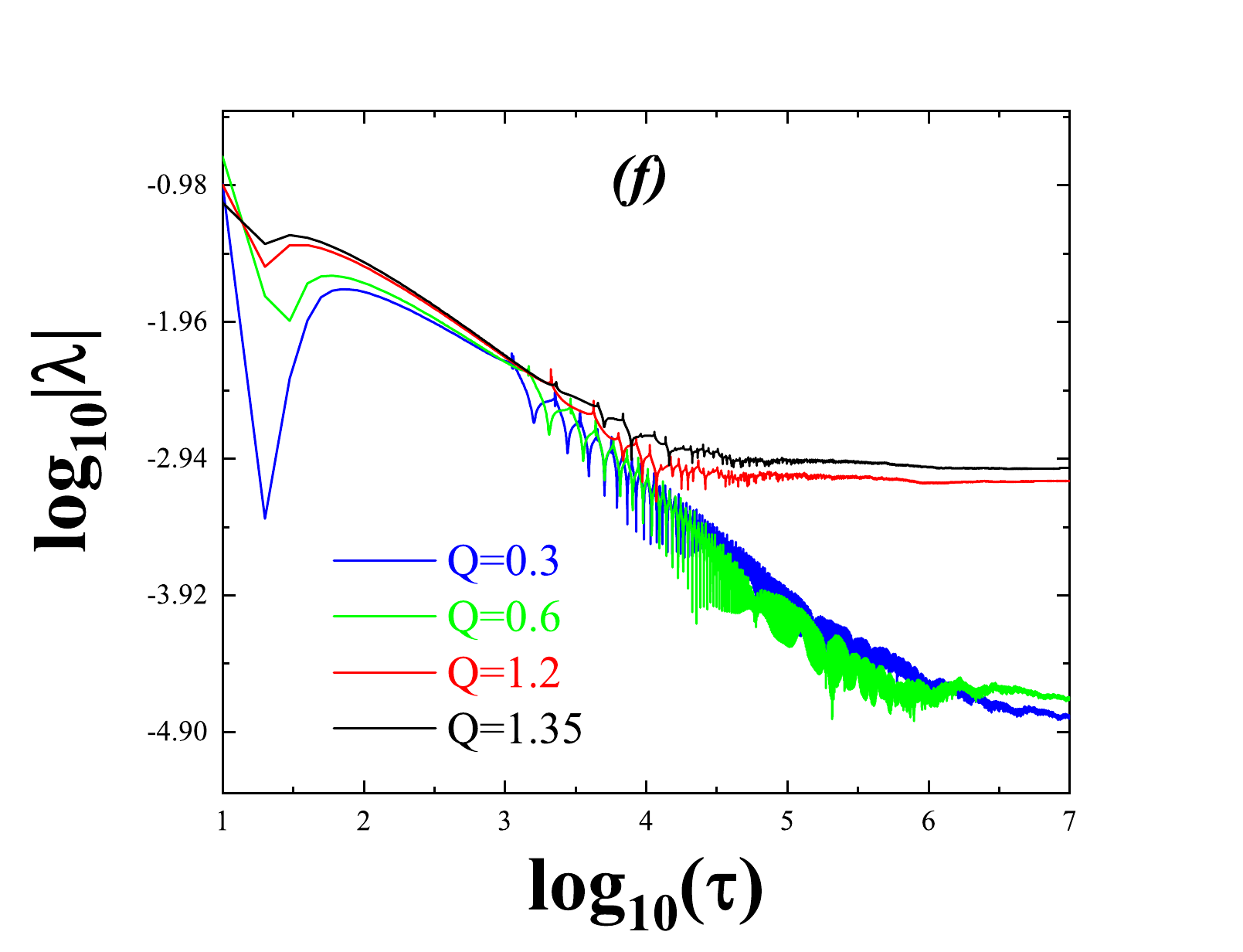}
		\phantomcaption
		\label{fig:subfig1f}
	\end{subfigure}
	
	\vspace{10pt}
	\centering 
	\begin{subfigure}[t]{0.32\textwidth}
		\includegraphics[width=\linewidth]{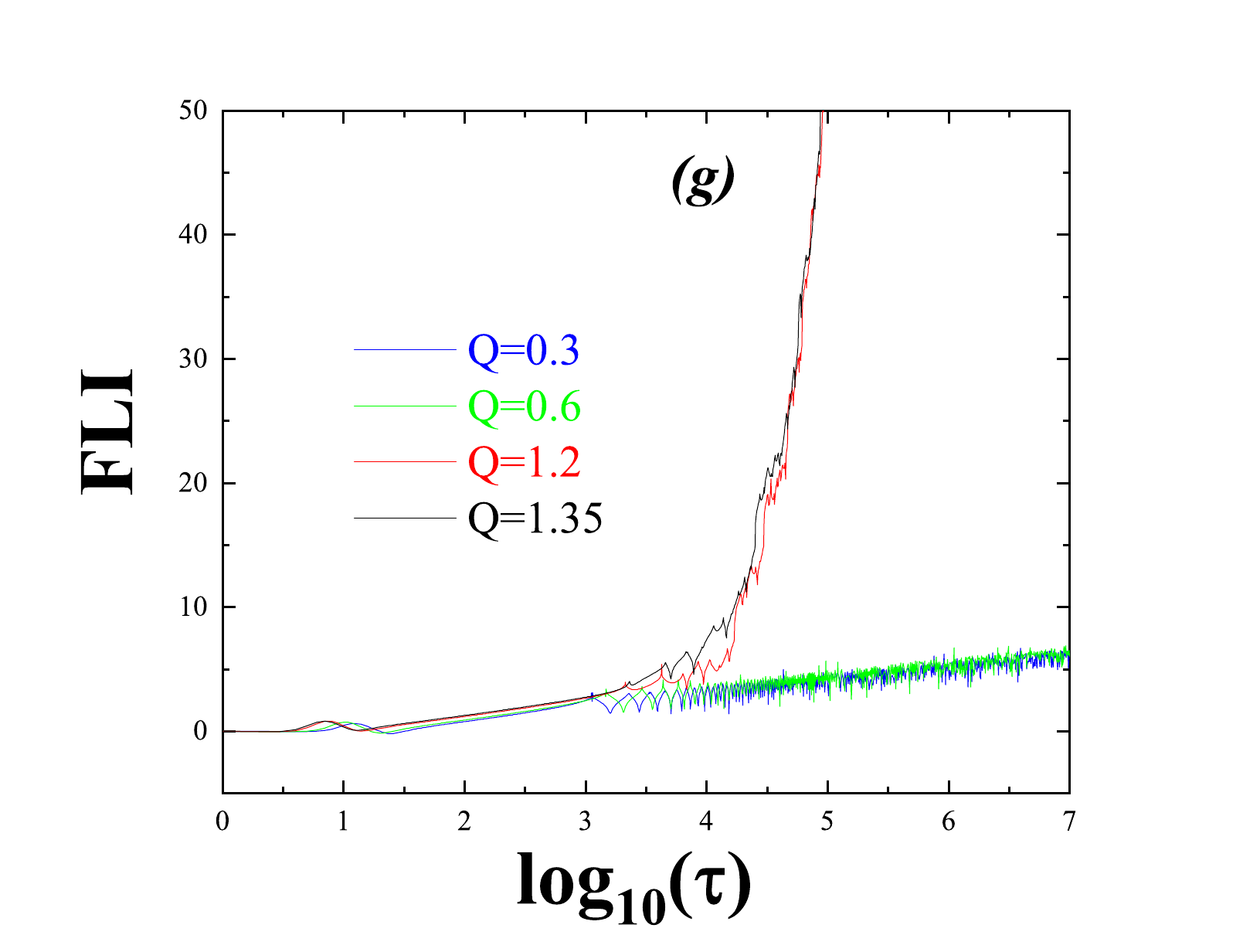}
		\phantomcaption
		\label{fig:subfig1g}
	\end{subfigure}
	
	\caption{\label{fig:1} Four chaos indicators for charged particle dynamics under black hole charge $Q$. The integration time is $\tau=10^7$. \textbf{(a)} Shannon entropy for four orbits with initial radius $r=6$, using parameters $E=0.98$, $L=2$, and $Q^*=qQ=-\sqrt{2} Q$ at four distinct $Q$ values. \textbf{(b)-(e)} Poincaré sections with identical parameters to \textbf{(a)}, showing different black hole charges: \textbf{(b)} $Q=0.3$, \textbf{(c)} $Q=0.6$, \textbf{(d)} $Q=1.2$, \textbf{(e)} $Q=1.35$. \textbf{(f)} MLEs for the four orbits. \textbf{(g)} FLIs for the four orbits.}
\end{figure*}

In this section, we first compute Shannon entropy for charged test particles orbiting a charged GMGHS black hole under varying initial conditions. Fixing parameters $E = 0.98$, $L = 2$, $Q^*=-\sqrt{2 }Q$, and step size $h = 1$. The initial conditions are selected as $r=6$, $\theta=\pi/2$, $p_r=0$, with $p_{\theta}$ derived from Eq. \eqref{8}. The integration time is $\tau=10^7$. Fig. \ref{fig:subfig1a} clearly shows that the entropy fluctuations are relatively large for the two trajectories at $Q=1.2$ and $Q=1.35$, while relatively small for those at $Q=0.3$ and $Q=0.6$. This indicates that the system is in a chaotic state when $Q=1.2$ and $Q=1.35$, whereas it exhibits ordered behavior when $Q=0.3$ and $Q=0.6$. 

Subsequently, we use Poincaré sections to verify the above conclusion. Figs. \ref{fig:subfig1b}-\ref{fig:subfig1e} depict the dynamic behavior of different parameter values $Q=0.3$, $Q=0.6$, $Q=1.2$, and $Q=1.35$, respectively. For each scenario, four orbital radii $r=5,6,10$, and $50$ are selected, with corresponding orbits labeled 1-4. Figs. \ref{fig:subfig1b}-\ref{fig:subfig1c} show that all four trajectories are closed orbits, indicating that the system is ordered. In Fig. \ref{fig:subfig1d}, orbits 3 and 4 remain closed (ordered system), while orbit 1 is characterized by randomly scattered points, signifying a transition to a chaotic state. Furthermore, orbit 2 displays disordered point distribution, confirming chaotic behavior. Finally, Fig. \ref{fig:subfig1e} clearly shows that orbits 1 and 2 are chaotic, while orbits 3 and 4 remain ordered. In conclusion, as the parameter $Q$ increases, the system transitions from an ordered state to a chaotic state. Under identical conditions, particles closer to the black hole are more susceptible to influence and exhibit chaotic behavior earlier than particles farther away. In addition, these conclusions can also be confirmed by MLE.

MLE is defined in \cite{Wu:2003pe} as $\lambda=\lim_{\tau\to\infty}\frac{1}{\tau}\ln\frac{d(\tau)}{d(0)}$, where $d (\tau)$ is the distance between two adjacent orbits at time $\tau$, and $d(0)$ is the initial distance between the two orbits. In Fig. \ref{fig:subfig1f}, it can be seen that when $Q=0.3$ and $Q=0.6$, the $\lambda$ values corresponding to the two orbits tend to zero, indicating ordered dynamics. However, when $Q = 1.2$ and $Q = 1.35$, the graph curves tend to be stable for a certain period of time, indicating that the system is in a chaotic state at this time.

Finally, a faster method to distinguish between order and chaos is FLI proposed by Froeschlé and Lega \cite{2000CeMDA..78..167F}. This indicator is redefined as $FLI = \log_{10} \frac{d(\tau)}{d(0)}$ based on two adjacent orbit methods \cite{Wu:2006rx}. Under the same initial conditions as in Fig. \ref{fig:subfig1a}, the FLI curves for $Q=1.2$ and $Q=1.35$ in Fig. \ref{fig:subfig1g} show exponential growth over time, indicating that the system is in a chaotic state. However, the graph curves of the two orbits at $Q=0.3$ and $Q=0.6$ show linear growth over time, indicating that the dynamic system is ordered at this time.

In the previous analysis, we first fix the parameter $Q^*$ and systematically investigate the influence of the parameter $Q$ on the chaotic dynamics of charged particles near a charged GMGHS black hole. Results from multiple analytical indicators consistently demonstrate that as the parameter $Q$ increases, the system transitions from an ordered state to a chaotic state. This suggests that the Shannon entropy, as a chaos indicator, can also be employed to study charged black holes in string theory. As shown in Fig. \ref{fig:subfig1a}, entropy fluctuations emerge when the system becomes chaotic, aligning with the findings in \cite{Cao:2024bjk}. Hence, entropy fluctuations can be regarded as a universal property of chaotic systems.

\begin{figure}[htbp]
	\centering
	\captionsetup[subfigure]{labelformat=empty} 
	
	\begin{subfigure}[t]{0.32\textwidth}
		\includegraphics[width=\linewidth]{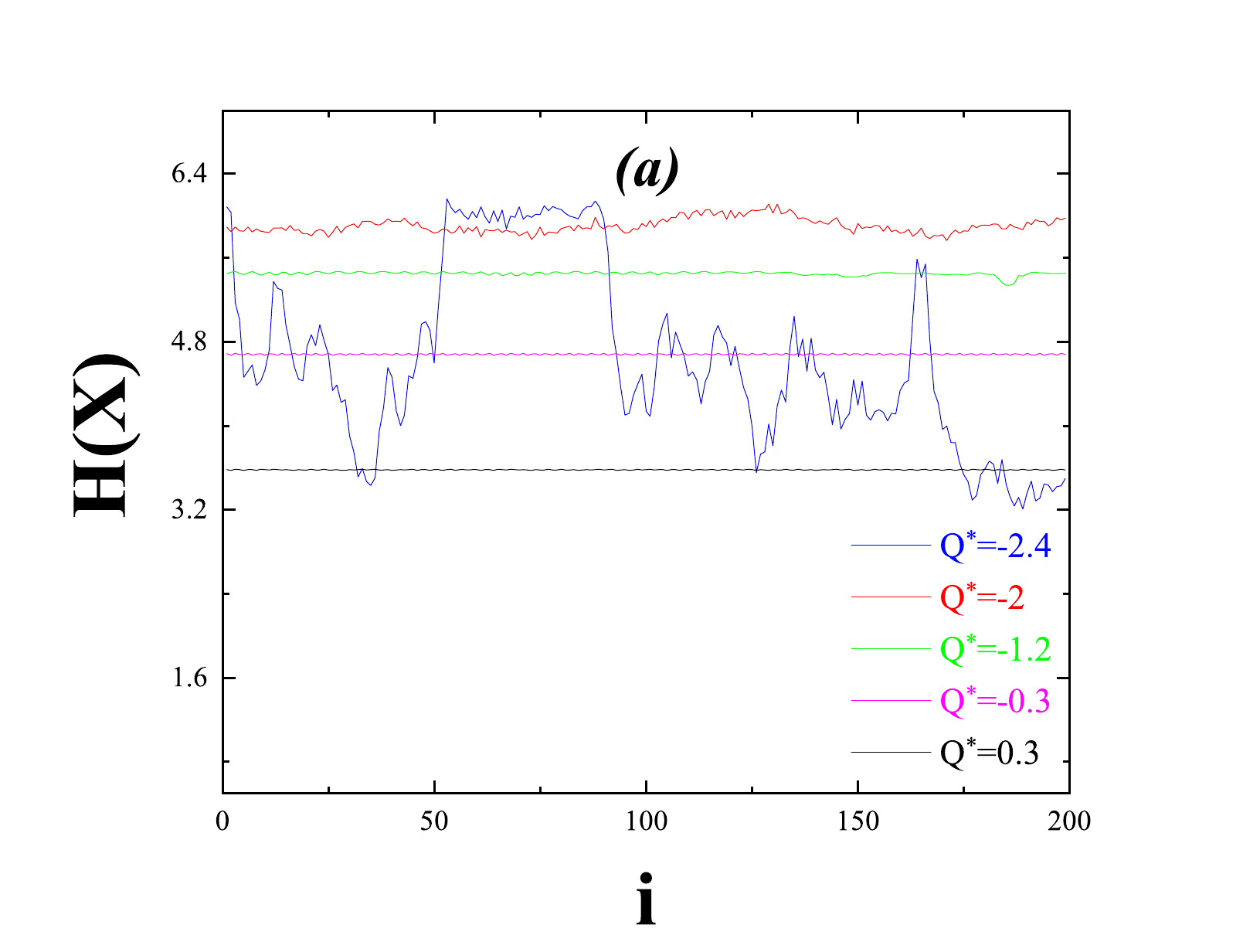}
		\phantomcaption 
		\label{fig:subfig2a}
	\end{subfigure}
	\hfill
	\begin{subfigure}[t]{0.32\textwidth}
		\includegraphics[width=\linewidth]{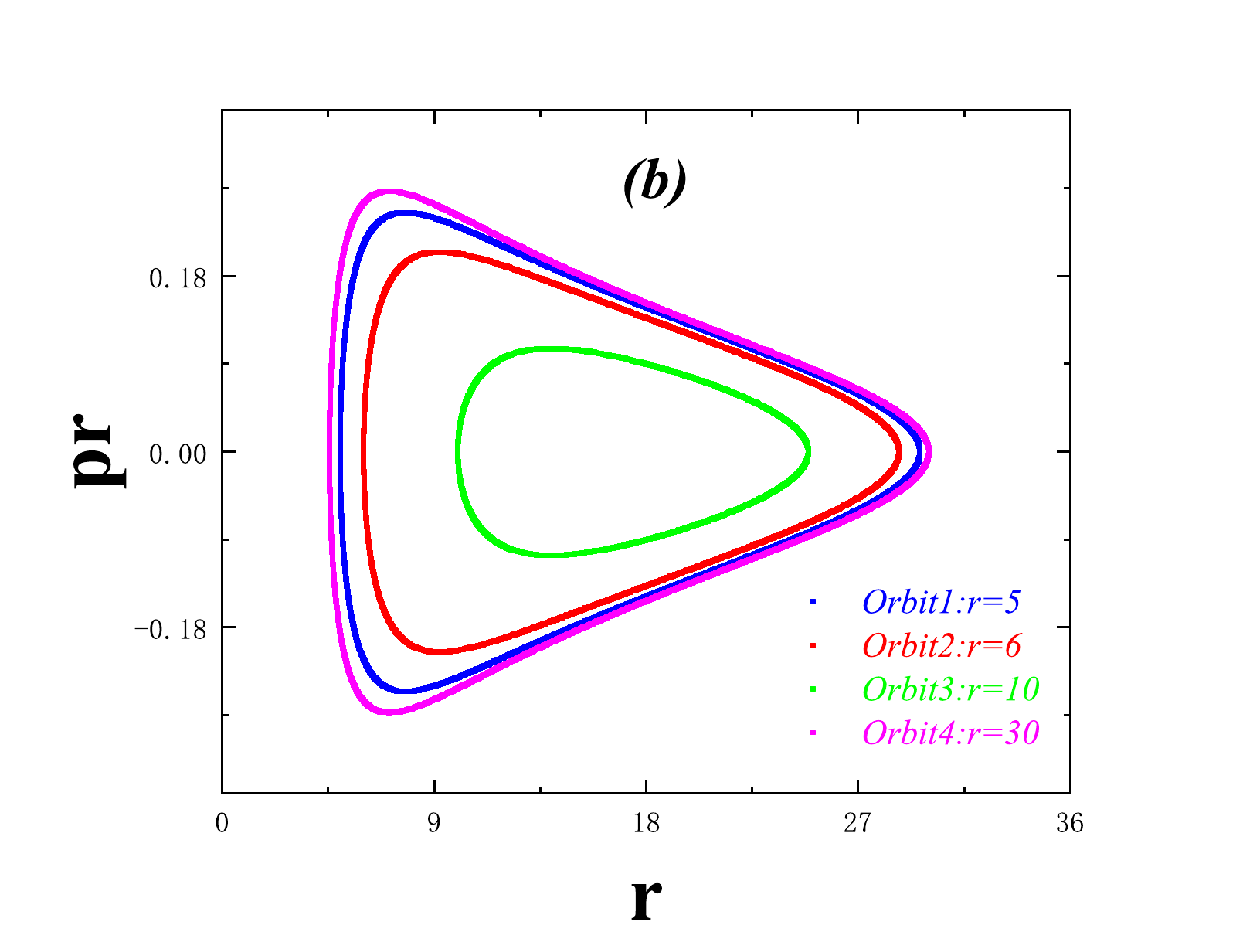}
		\phantomcaption
		\label{fig:subfig2b}
	\end{subfigure}
	\hfill
	\begin{subfigure}[t]{0.32\textwidth}
		\includegraphics[width=\linewidth]{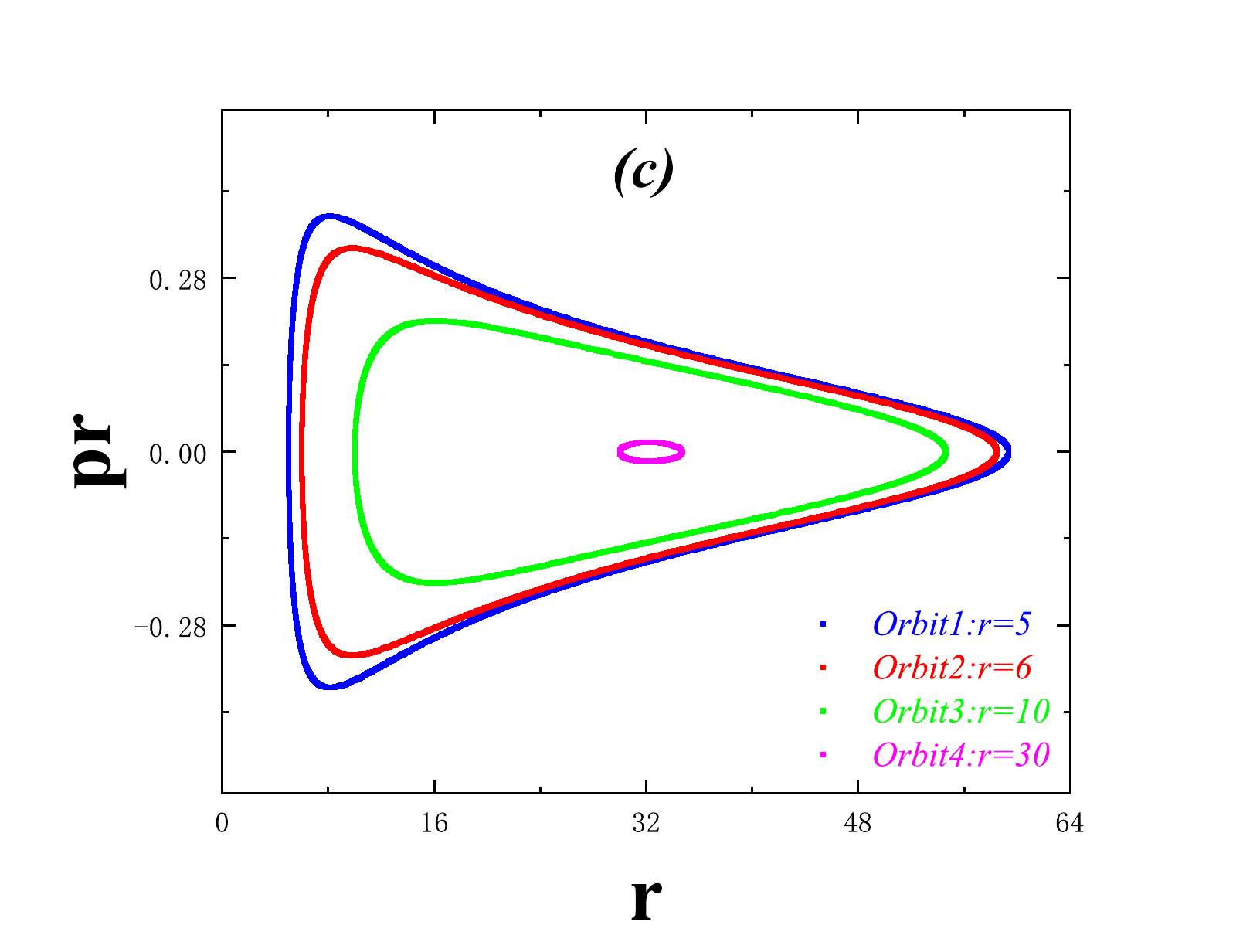}
		\phantomcaption
		\label{fig:subfig2c}
	\end{subfigure}
	
	\vspace{10pt} 
	\begin{subfigure}[t]{0.32\textwidth}
		\includegraphics[width=\linewidth]{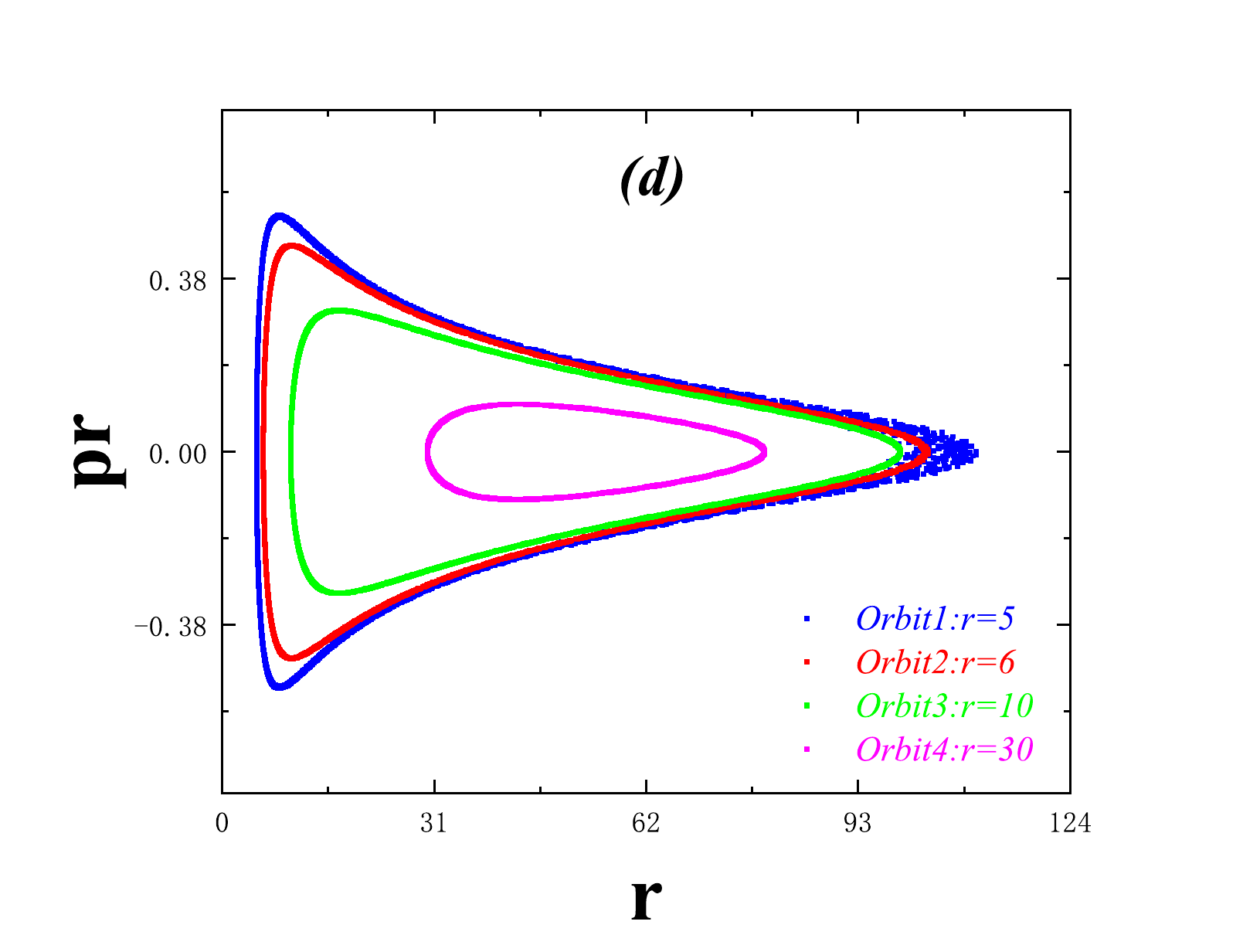}
		\phantomcaption
		\label{fig:subfig2d}
	\end{subfigure}
	\hfill
	\begin{subfigure}[t]{0.32\textwidth}
		\includegraphics[width=\linewidth]{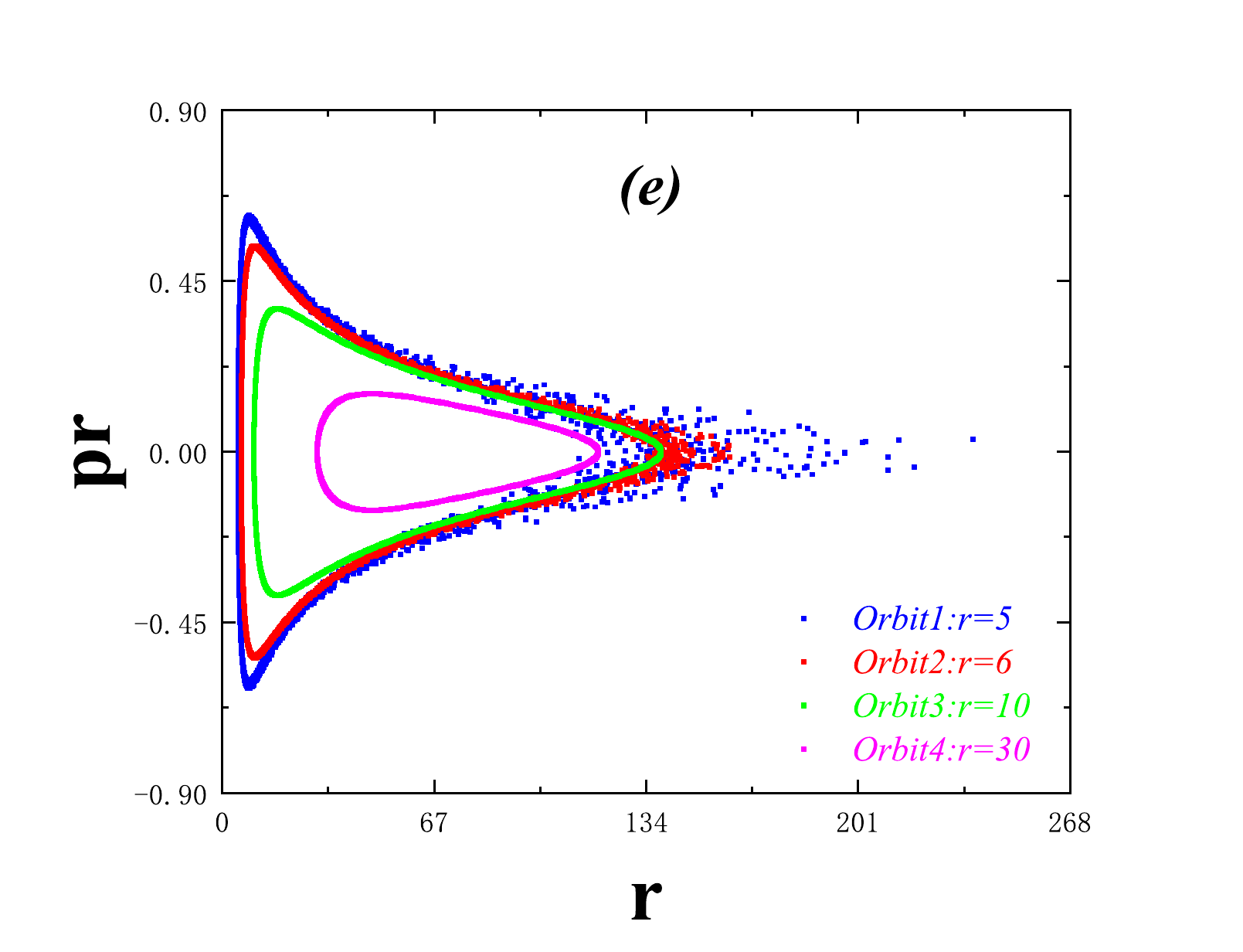}
		\phantomcaption
		\label{fig:subfig2e}
	\end{subfigure}
	\hfill
	\begin{subfigure}[t]{0.32\textwidth}
		\includegraphics[width=\linewidth]{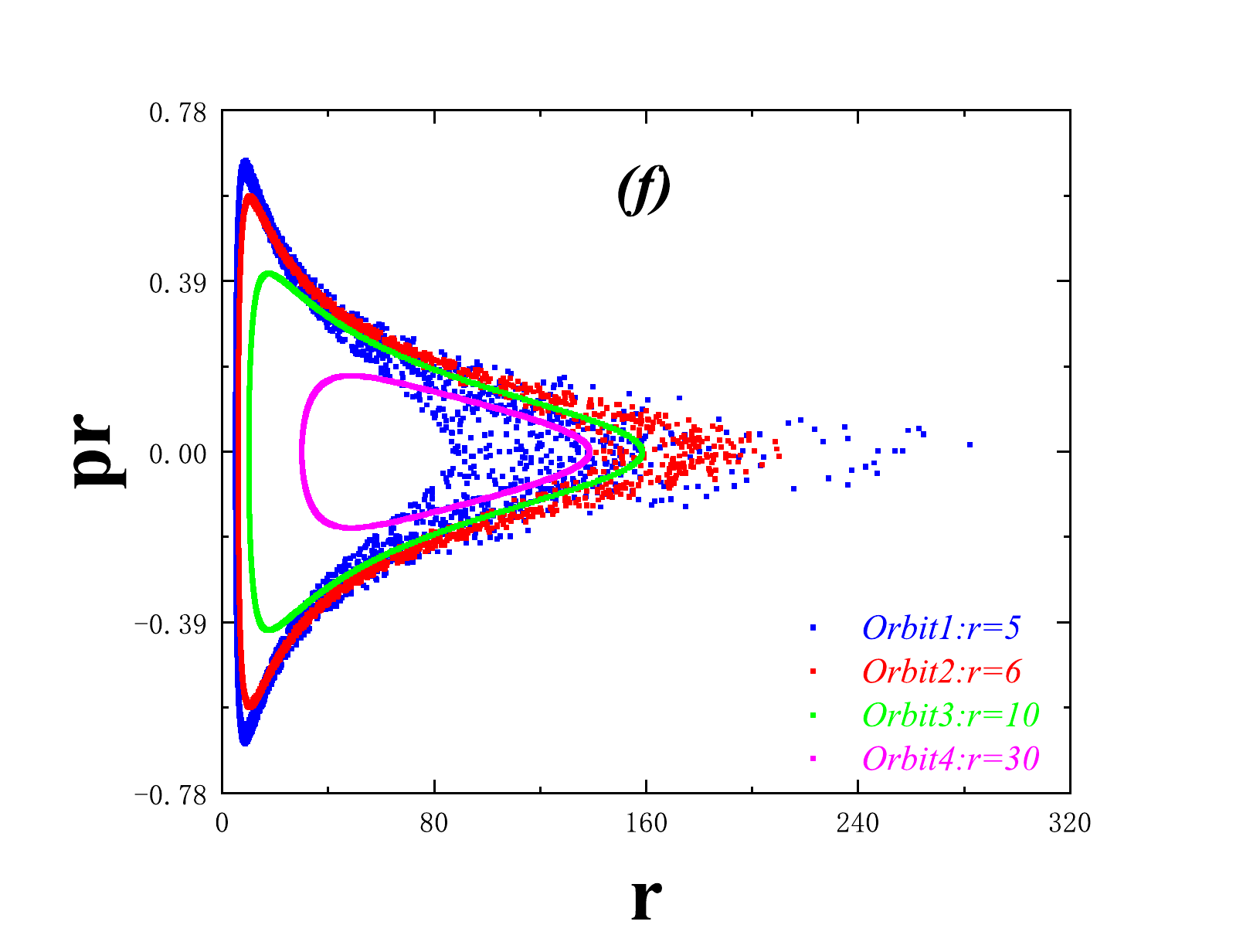}
		\phantomcaption
		\label{fig:subfig2f}
	\end{subfigure}
	
	\vspace{10pt}
	\begin{subfigure}[t]{0.32\textwidth}
		\includegraphics[width=\linewidth]{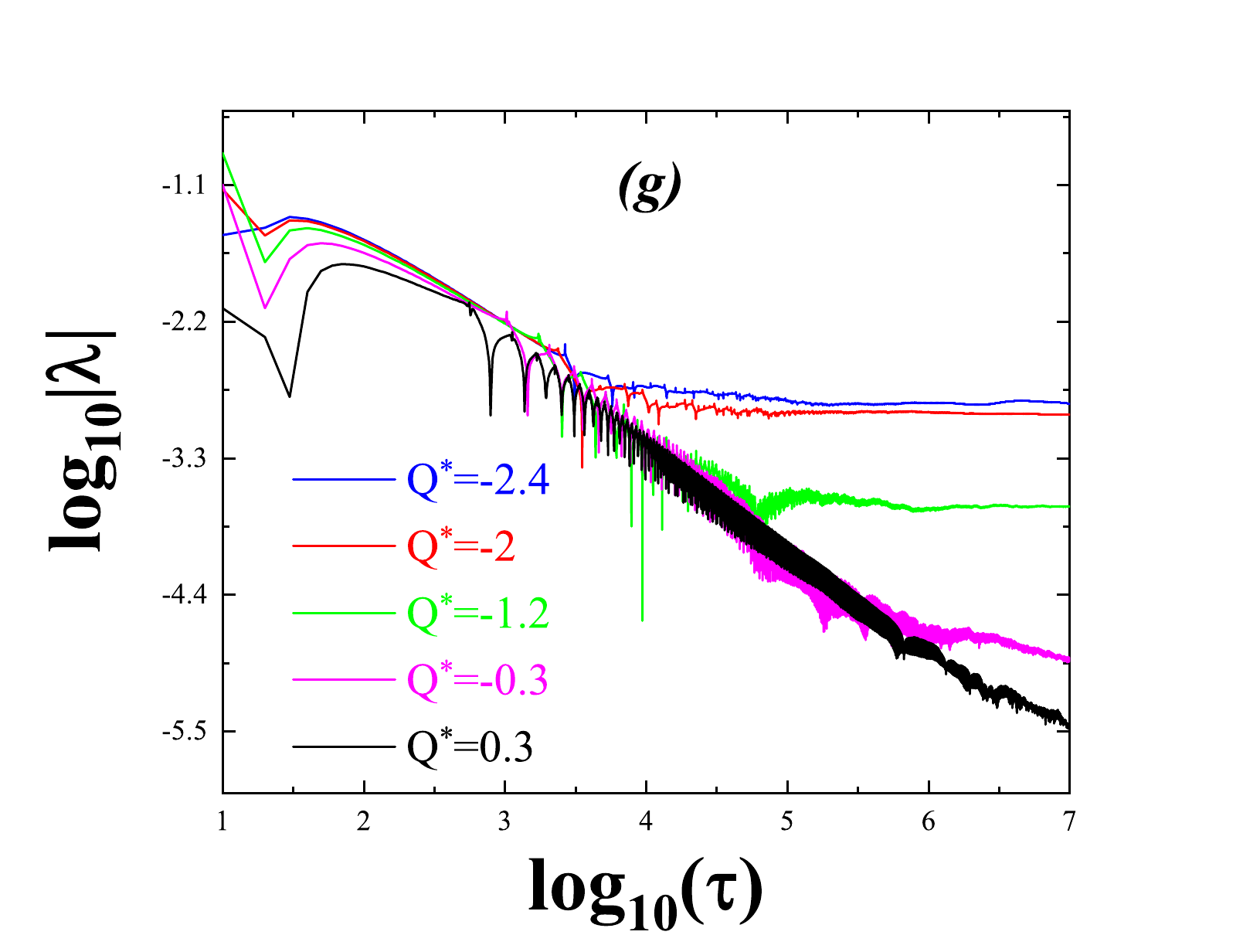}
		\phantomcaption
		\label{fig:subfig2g}
	\end{subfigure}
	\begin{subfigure}[t]{0.32\textwidth}
		\includegraphics[width=\linewidth]{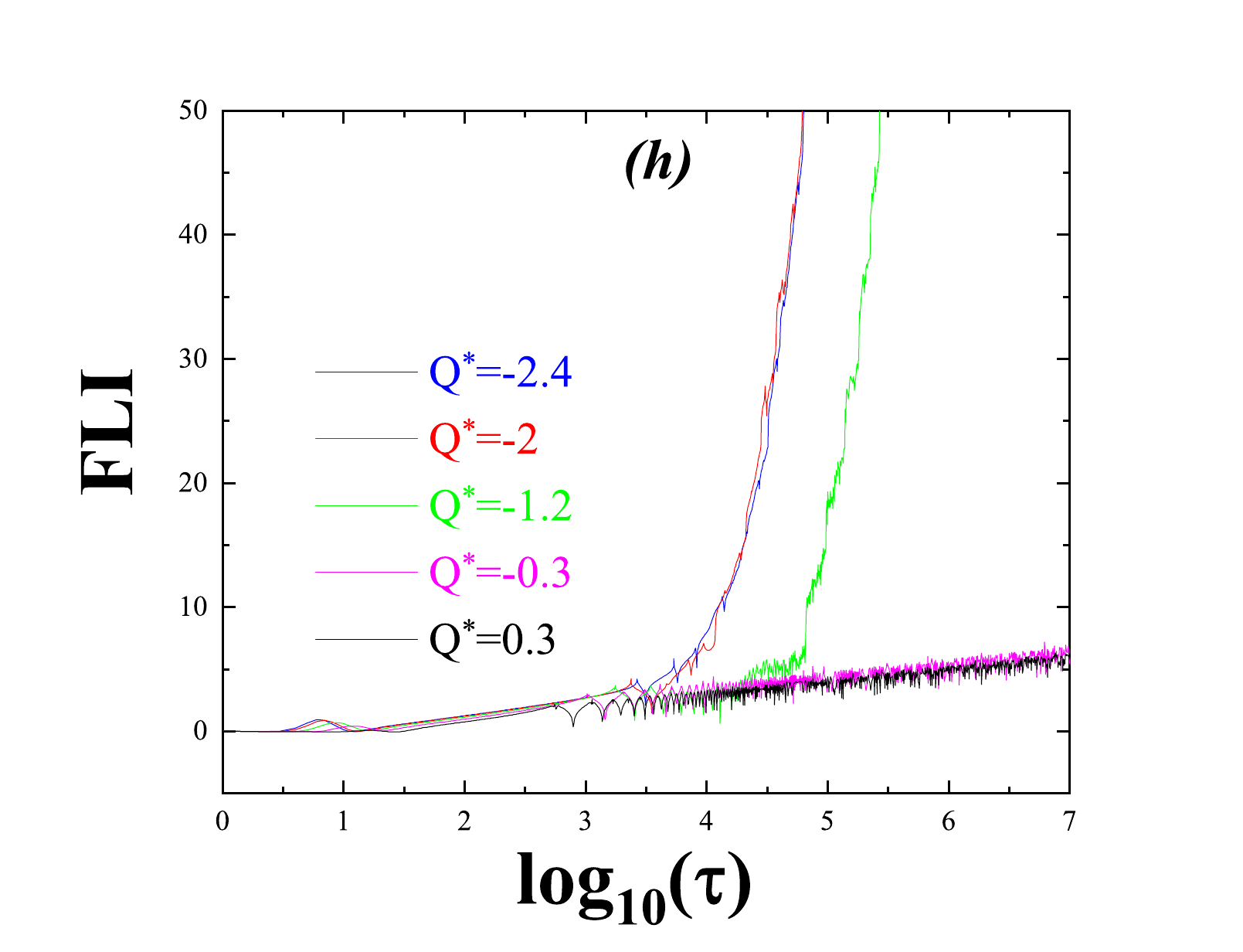}
		\phantomcaption
		\label{fig:subfig2h}
	\end{subfigure}
	
	\caption{\label{fig:2} Influence of parameter $Q^*$ on the chaotic dynamics of a charged test particle. Parameters are set as in Fig. \ref{fig:1}, but with $Q$ fixed at 1.2. \textbf{(a)} Shannon entropy for five orbits. \textbf{(b)-(f)} Poincaré sections corresponding to different $Q^*$ values: \textbf{(b)} $Q^*=0.3$, \textbf{(c)} $Q^*=-0.3$, \textbf{(d)} $Q^*=-1.2$, \textbf{(e)} $Q^*=-2$, \textbf{(f)} $Q^*=-2.4$. \textbf{(g)} MLEs and \textbf{(h)} FLIs for the five orbits.}
\end{figure}

Subsequently, we fix $Q=1.2$ to explore the influence of the parameter $Q^*$. Similar to the parameter settings in Fig. \ref{fig:subfig1a}, Fig. \ref{fig:subfig2a} presents the temporal evolution of the Shannon entropy. All orbits exhibit entropy fluctuations. For $Q^*= -2.4$, $Q^*= -2$, and $Q^*= -1.2$, the fluctuation intensity decreases with increasing Q*, being strongest at $Q^*= -2.4$ and relatively weakest at $Q^*= -1.2$. In contrast, orbits with $Q^*= -0.3$ and $Q^*= 0.3$ exhibit significantly smaller fluctuation amplitudes. Then, using the same initial conditions as Figs. \ref{fig:subfig1b}-\ref{fig:subfig1e}, we plot the Poincaré sections for orbits at radii $r=5,6,10$, and $30$ under five scenarios: $Q^*=0.3, -0.3, -1.2, -2, -2.4$, as shown in Figs. \ref{fig:subfig2b}-\ref{fig:subfig2f}. At $Q^*=0.3$ or -0.3, all four trajectories in Figs. \ref{fig:subfig2b} and \ref{fig:subfig2c} exhibit highly regular ring structures. This indicates regular and integrable particle motion where chaotic behavior is strongly suppressed. As $Q^*$ decreases, enhancing the Coulomb attraction, dynamical complexity emerges at $Q^*=-1.2$. Fig. \ref{fig:subfig2d} shows that the $r=5$ trajectory begins to exhibit chaos. However, the section structures for $r=6$, $r=10$, and $r=30$ remain regular. When $Q^*=-2$, pronounced chaotic features in the $r=5$ trajectory are revealed in Fig. \ref{fig:subfig2e}. It is noteworthy that while chaotic regions exist in the $r=6$ trajectory, stable periodic orbits are also partially preserved. As the orbital radius increases to $r=10$ and $r=30$, chaotic behavior diminishes rapidly and the section structures regain high regularity. In Fig. \ref{fig:subfig2f}, at $Q^*=-2.4$, the $r=5$ trajectory still exhibits strong chaotic characteristics. For the $r=6$ trajectory, chaos remains significant, with observable expansion of chaotic regions compared to $Q^*=-2$, leaving limited ordered structures. Conversely, particle motion for $r=10$ and $r=30$ returns to a highly regular state. Results from Figs. \ref{fig:subfig2b}-\ref{fig:subfig2f} reveal that trajectories at $r=10$ and $r=30$ remain consistently ordered. The trajectories at $r=5$ and $r=6$ are ordered when $Q^*=0.3$ and $Q^*=-0.3$, respectively. However, these trajectories become chaotic at $Q^*=-1.2$, $-2$, and $-2.4$. We therefore conclude that decreasing $Q^*$ drives a transition from order to chaos, and particles near the black hole are more susceptible to this effect. To corroborate these findings, the evolution of MLE versus $\log_{10} (\tau)$ for different parameter values $Q^*$ is shown in Fig. \ref{fig:subfig2g}. The confirmation from the results shows that chaotic behavior occurs at $Q^* = -2.4, -2$, and $-1.2$, while regular behavior occurs at $Q^* = -0.3$ and $0.3$. Similarly, Fig. \ref{fig:subfig2h} depicts the evolution of the FLIs with respect to $\log_{10} (\tau)$. For $Q^*=-2.4, -2$, and $Q^*= -1.2$, the FLIs exhibits exponential growth, signifying the onset of chaotic motion. In contrast, for $Q^*=-0.3$ and 0.3, the FLIs shows linear growth, indicating a regular state. The results from FLIs are the same as MLE. In summary, the stability of charged particle motion in the external field of a GMGHS dilaton black hole is highly sensitive to the Coulomb parameter $Q^*$. Under strong Coulomb attraction, the system becomes chaotic, whereas it remains ordered under conditions of weak Coulomb attraction or Coulomb repulsion.

\section{The magnetically charged black hole} \label{section4}

In Section \ref{section3}, we employed four chaos indicators to analyze the dynamical behavior and chaotic properties of charged test particles around electrically charged GMGHS dilaton black holes. The results demonstrate that increasing the black hole charge $Q$ or decreasing the parameter $Q^*$ drives a system transition from order to chaos. We now apply these same indicators to investigate chaotic dynamics near magnetically charged GMGHS black holes. 

For a magnetically charged black hole with magnetic charge $Q_m$, where the magnetic field originates from the vector potential component $A_{\varphi}=-Q_{m}\cos\theta$, the corresponding Lagrangian is
\begin{equation}\label{16}
	\mathcal{L}=\frac{1}{2}\left[-f\dot{t}^2+\frac{\dot{r}^2}{f}+p^2(\dot{\theta}^2+\sin^2\theta\dot{\varphi}^2)\right]-qQ_m\cos\theta\dot{\varphi}.
\end{equation}
The conserved quantities are energy $\tilde{E}$ and angular momentum $\tilde{L}$:
\begin{eqnarray}\label{17}
	-{{\tilde{E}}} &=& {p_t} = -f\dot{t},\nonumber\\
	{{\tilde{L}}} &=& {p_\varphi } =p^2\sin^2\theta\dot{\varphi}-qQ_m\cos\theta.
\end{eqnarray}
The Hamiltonian governing charged particle motion is
\begin{eqnarray}\label{18}
	K&=&p_{\mu}\dot{x}^{\mu}-\mathcal{L}=\frac{1}{2}g^{\mu\nu}(p_{\mu}-qA_{\mu})(p_{\nu}-qA_{\nu}) \nonumber\\
	&=&\frac{1}{2}\left[\frac{\tilde{E}^{2}r}{2-r}
	+\frac{\csc^{2}\theta(\tilde{L}+qQ_m\cos\theta)^{2}}{r^{2}-rQ_m^{2}}\right]+\frac{(r-2)p_r^{2}}{2r}\nonumber\\
	&&+\frac{p_\theta^{2}}{2r^{2}-2rQ_m^{2}}. 
\end{eqnarray}
Similar to Eq. \eqref{9}, the Hamiltonian satisfies $K = -\frac{1}{2}$ and is decomposed into four integrable subsystems:
\begin{equation}\label{19}
	{K} = {K_1} + {K_2} + {K_3} + {K_4}.
\end{equation}
where
\begin{eqnarray}\label{20}
	{K_1} &=& \frac{1}{2}\left[\frac{\tilde{E}^{2}r}{2-r}
	+\frac{\csc^{2}\theta(\tilde{L}+qQ_m\cos\theta)^{2}}{r^{2}-rQ_m^{2}}\right],\nonumber \\
	{K_2} &=& \frac{p_r^2}{{2}},\nonumber \\
	{K_3}&=& \frac{{-p_r^2}}{r},\nonumber \\
	{K_4} &=& \frac{p_\theta^{2}}{2r(-Q_m^{2}+r)}.
\end{eqnarray}
The analytical solution operators for these subsystems in Eq. \eqref{20} are designated as ${{{\cal K}}_1}$, ${{{\cal K}}_2}$, ${{{\cal K}}_3}$, and ${{{\cal K}}_4}$. Then, by replacing the operators ${{{\cal H}}_1}$, ${{{\cal H}}_2}$, ${{{\cal H}}_3}$, and ${{{\cal H}}_4}$ in Eqs. \eqref{12} and \eqref{13} with the analytic solutions for the magnetized black hole case, the corresponding $PR{K_6}4$ algorithm can be constructed according to Eq. \eqref{14}. Subsequently, we apply the $PR{K_6}4$ algorithm and four chaos indicators to analyze the chaotic dynamics of charged particles around a magnetized black hole. Since $\theta=\pi/2$ is adopted in this work, the term $qQ_m\cos\theta$ vanishes in Eq. \eqref{18}. Therefore, we temporarily neglect the interaction between the magnetic charge $Q_m$ and particle charge $q$, focusing solely on the influence of the parameter $Q_m$.

\begin{figure*}[htbp]
	\centering
	\captionsetup[subfigure]{labelformat=empty} 
	\begin{subfigure}[t]{0.32\textwidth}
		\includegraphics[width=\linewidth]{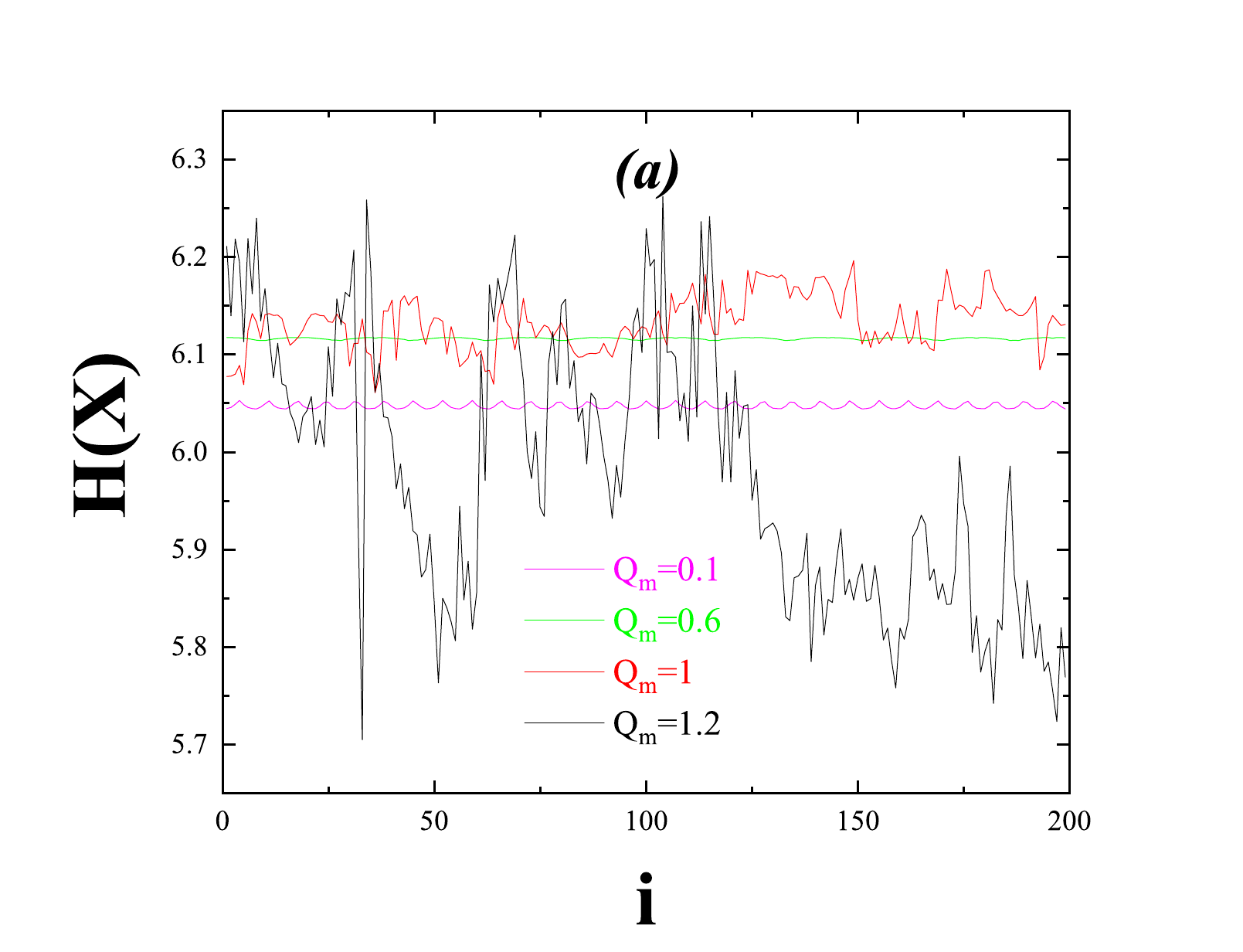}
		\phantomcaption 
		\label{fig:subfig3a}
	\end{subfigure}
	\hfill
	\begin{subfigure}[t]{0.32\textwidth}
		\includegraphics[width=\linewidth]{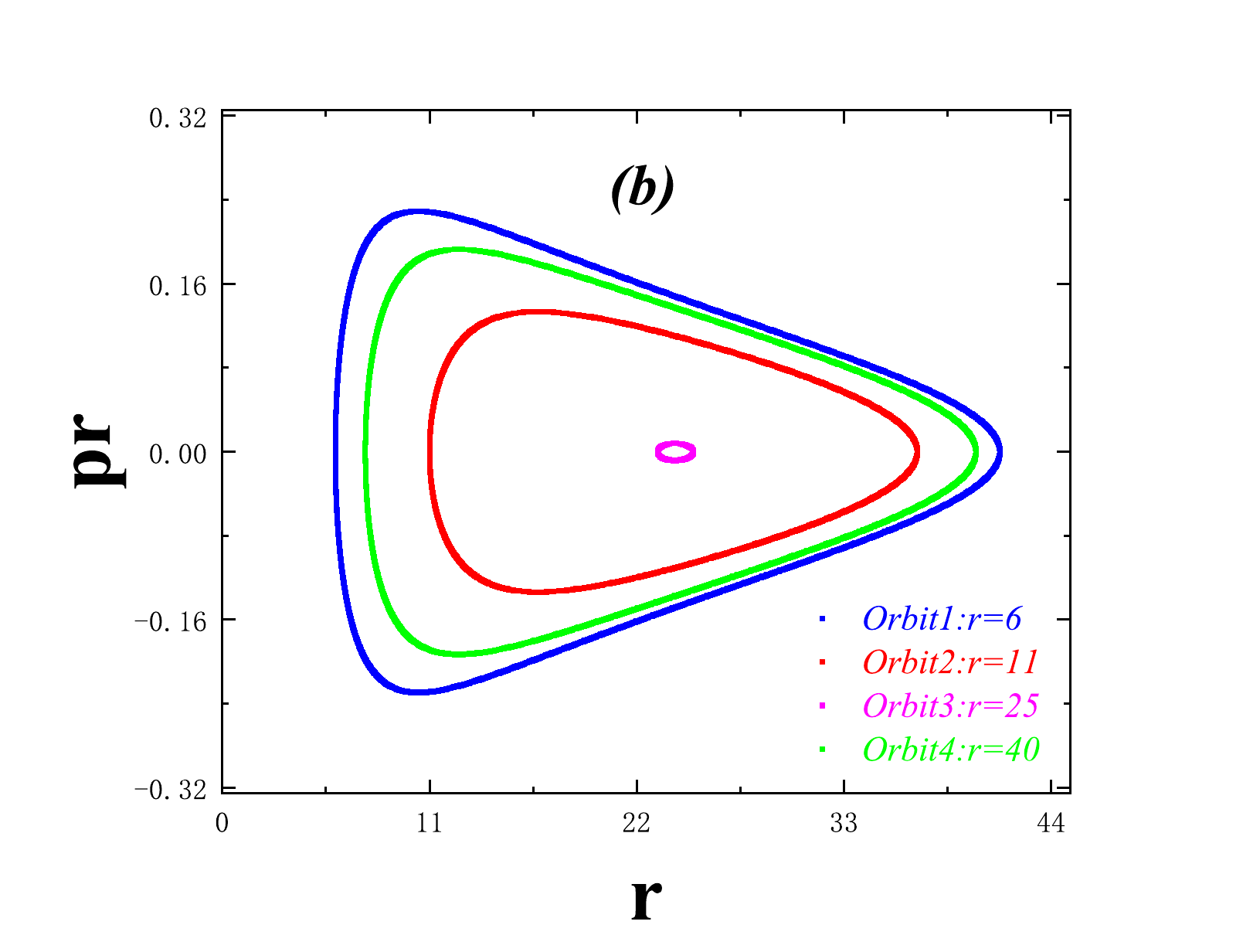}
		\phantomcaption
		\label{fig:subfig3b}
	\end{subfigure}
	\hfill
	\begin{subfigure}[t]{0.32\textwidth}
		\includegraphics[width=\linewidth]{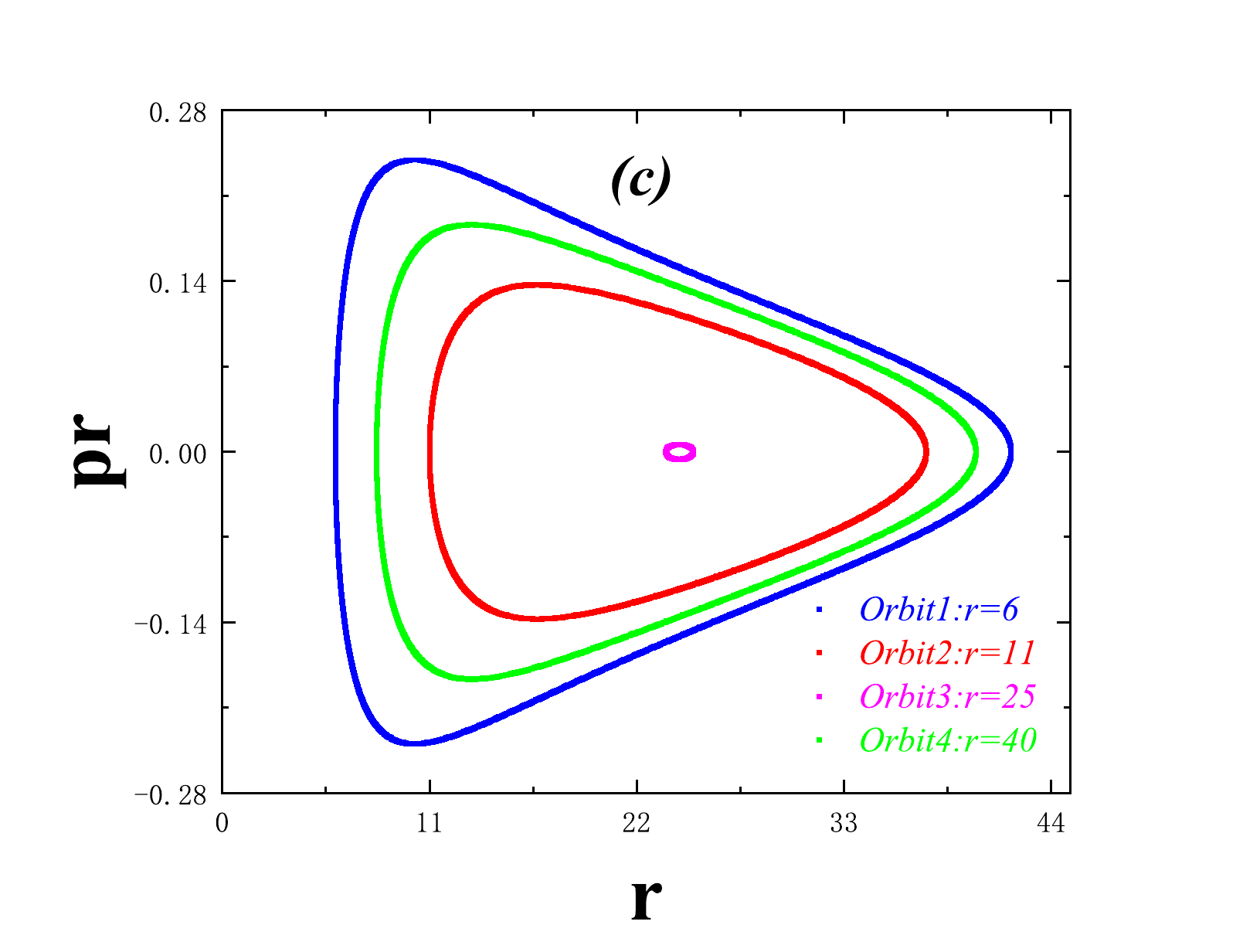}
		\phantomcaption
		\label{fig:subfig3c}
	\end{subfigure}
	
	\vspace{10pt} 
	\begin{subfigure}[t]{0.32\textwidth}
		\includegraphics[width=\linewidth]{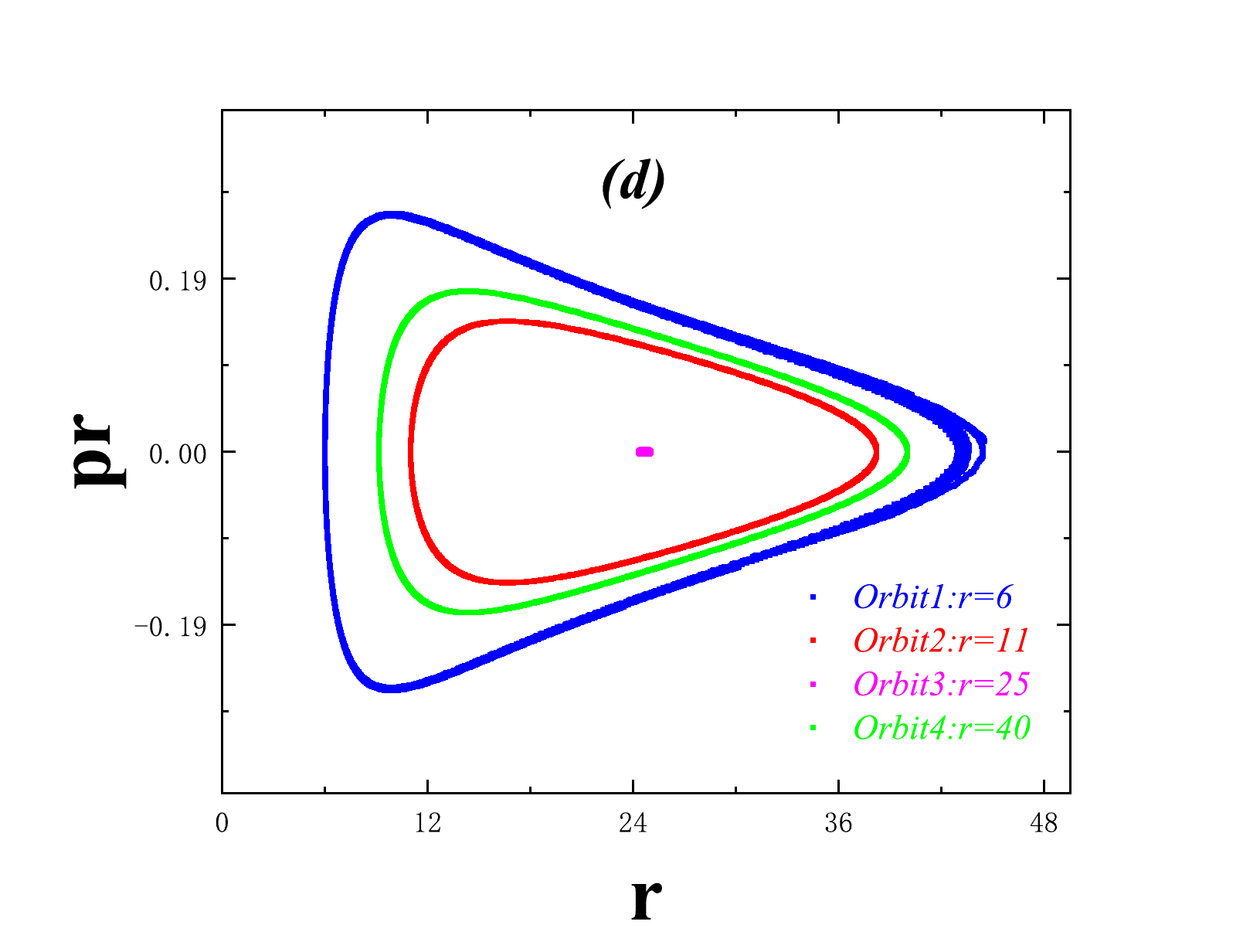}
		\phantomcaption
		\label{fig:subfig3d}
	\end{subfigure}
	\hfill
	\begin{subfigure}[t]{0.32\textwidth}
		\includegraphics[width=\linewidth]{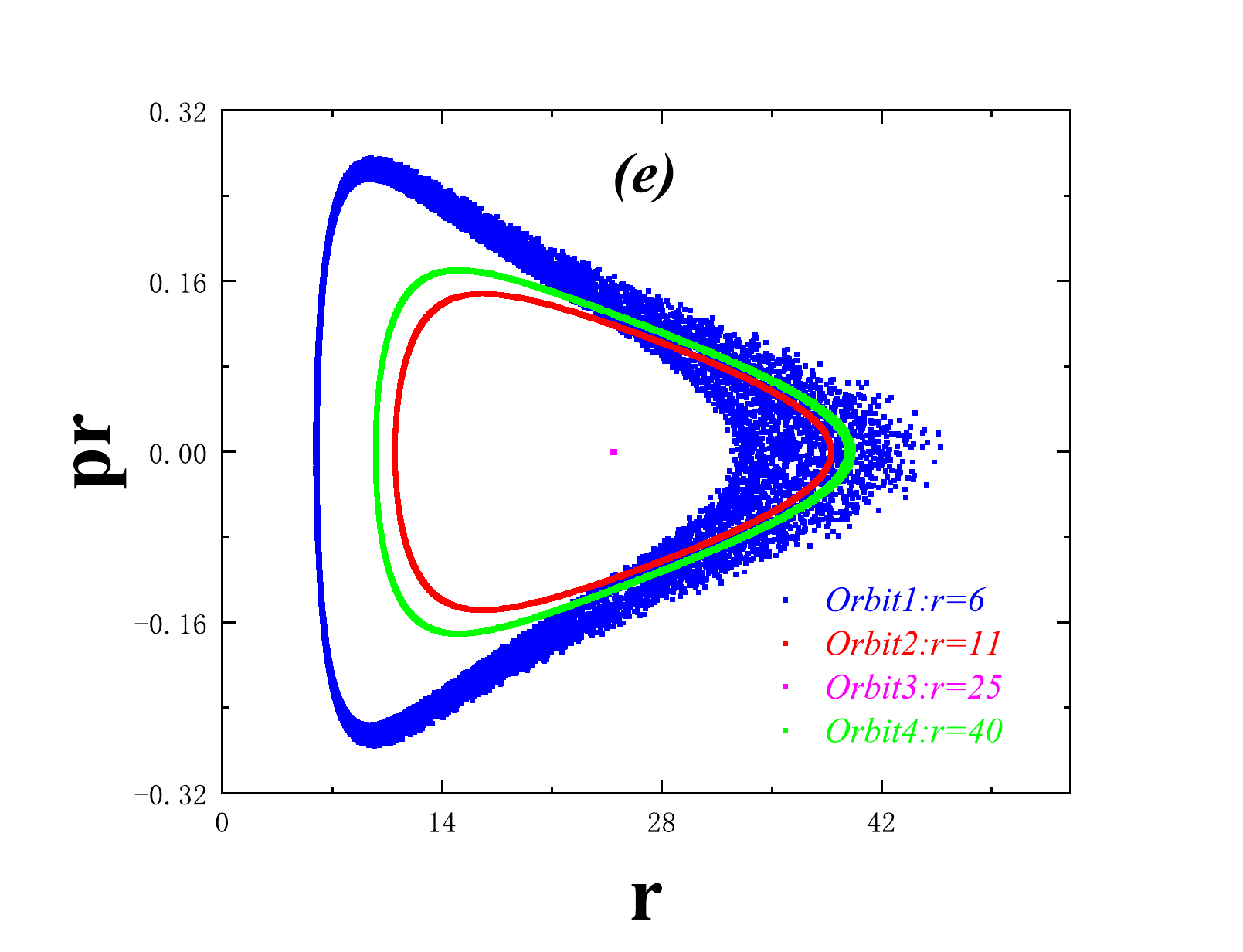}
		\phantomcaption
		\label{fig:subfig3e}
	\end{subfigure}
	\hfill
	\begin{subfigure}[t]{0.32\textwidth}
		\includegraphics[width=\linewidth]{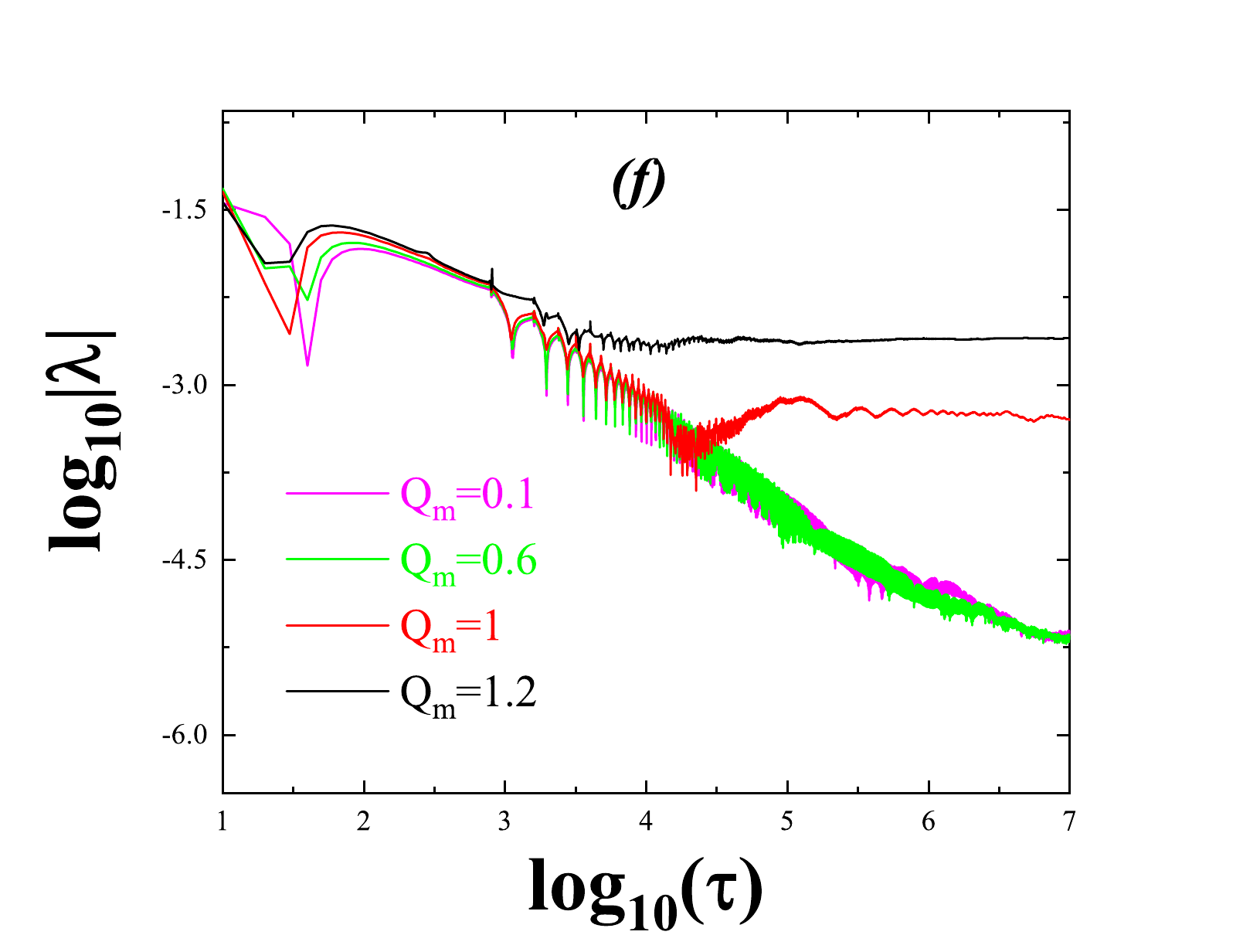}
		\phantomcaption
		\label{fig:subfig3f}
	\end{subfigure}
	
	\vspace{10pt}
	\begin{subfigure}[t]{0.32\textwidth}
		\centering 
		\includegraphics[width=\linewidth]{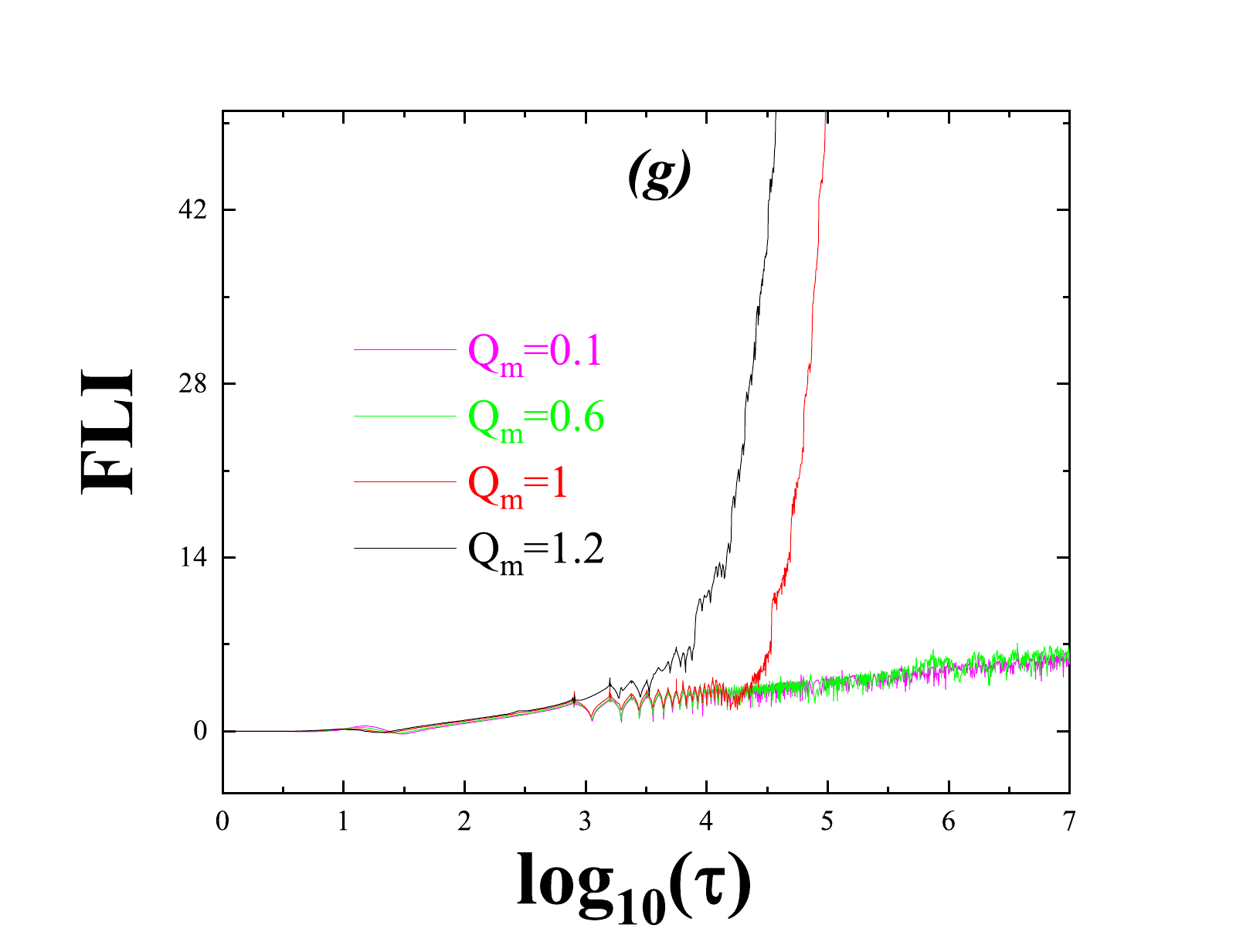}
		\phantomcaption
		\label{fig:subfig3g}
	\end{subfigure}
	
	\caption{Chaotic dynamics of a charged test particle near a black hole with magnetic charge $Q_m$. \textbf{(a)} Shannon entropy for four orbits at distinct $Q_m$, with $r=6$, $E=0.98$, $L=2$. \textbf{(b)-(e)} Poincaré sections with different magnetic charges $Q_m$: \textbf{(b)} $Q_m=0.1$, \textbf{(c)} $Q_m=0.6$, \textbf{(d)} $Q_m=1$, \textbf{(e)} $Q_m=1.2$. \textbf{(f)} MLEs and \textbf{(g)} FLIs for the four orbits.}
	\label{fig:3}
\end{figure*}

Fig. \ref{fig:subfig3a} clearly shows that Shannon entropy exhibits significant irregular fluctuations over time at $Q_m=1$ or 1.2. In contrast, its fluctuations are minimal and highly regular at $Q_m=0.1$ or 0.6. This indicates that chaos occurs at $Q_m=1$ or 1.2, while order is maintained at $Q_m=0.1$ or 0.6. Subsequently, Figs. \ref{fig:subfig3b}-\ref{fig:subfig3e} present Poincaré sections for charged particles at $Q_m=0.1, 0.6, 1$, and 1.2, respectively, analyzing orbits at $r=6$, $r=11$, $r=25$, and $r=40$. Specifically, Figs. \ref{fig:subfig3b} and \ref{fig:subfig3c} reveal all four orbits forming distinct, closed KAM tori, indicating orderly behavior at $Q_m=0.1$ and 0.6. In Fig. \ref{fig:subfig3d}, the orbit at $r=6$ consists of scattered points, while the orbits at $r=11,25$, and 40 remain closed curves. This indicates that chaotic behavior emerges at $r=6$. Fig. \ref{fig:subfig3e} further demonstrates dense irregular points at $r=6$, signifying strong chaos, whereas orbits at $r=11,25$, and 40 remain ordered. Thus, the system transitions from order to chaos as $Q_m$ increases. Similarly, the same initial conditions as Fig. \ref{fig:subfig1f} are adopted in Figs. \ref{fig:subfig3f}-\ref{fig:subfig3g}. Fig. \ref{fig:subfig3f} shows the time evolution of MLE for charged particle orbits in the GMGHS dilaton black hole. The four curves correspond to parameters $Q_m=0.1, 0.6, 1$, and 1.2. Subsequently, it is observed that the curves decrease continuously over time at $Q_m=0.1$ or 0.6, asymptotically approaching zero. However, the curves gradually become stable when $Q_m=1$ or 1.2. These results confirm that the system is orderly at $Q_m=0.1$ or $Q_m=0.6$, but chaotic at $Q_m=1$ or $Q_m=1.2$. In Fig. \ref{fig:subfig3g}, the linear  growth over time of the FLIs for $Q_m=0.1$ and $Q_m=0.6$ indicates order, whereas the exponential growth over time of the FLIs for $Q_m=1$ and $Q_m=1.2$ signifies chaos. Therefore, we can conclude that as $Q_m$ increases, the system gradually transitions from an ordered state to a chaotic state.

\section{Conclusions} \label{section5}
The GMGHS charged black hole, the solution emerges from the four-dimensional reduction of heterotic string theory via six-torus compactification. Because this solution carries charge, the trajectories of electrically charged probes exhibit distinct dynamics absent for neutral counterparts. This contrast stems from additional electrodynamic coupling between the probe and the background field. Thus, the present work investigates the chaotic dynamics of charged test particles around electrically and magnetically charged GMGHS black holes. Due to the significant advantages of the $\mathrm{PRK}_64$ algorithm in terms of stability and high accuracy during long-term numerical integration, we construct this optimized fourth-order explicit symplectic algorithm by decomposing the Hamiltonian into four integrable parts. Subsequently, employing four distinct chaos indicators, we systematically examine the influence of varying parameters $Q$, $Q^*$, and $Q_m$ on the chaotic dynamics of the charged particles. We find that as parameters $Q$ and $Q_m$ increase, the system transitions from an ordered state to a chaotic state. Conversely, a decrease in parameter $Q^*$ also induces a transition from order to chaos in the system.

\begin{acknowledgments}
	This work is supported by the Natural Science Foundation of China under Grant Nos. 12473074, 12073008, 11703005.
\end{acknowledgments}

\appendix
\section*{APPENDIX}\label{appendix}

\section{Discrete difference scheme of algorithm $PR{K_6}4$}
The equations of motion for the sub-Hamiltonians in Eq. \eqref{11} are given as following,
\begin{eqnarray}
	{{{\cal H}}_1}:\frac{d{p_r}}{{d\tau}} &=&-\frac{\partial H_{1}}{\partial r}=-\frac{(-Q^*+Er)(Q^*+Er-Q^*r)}{r^2(-2+r)^2}+\frac{L^{2}\csc^{2}\theta(-Q^2+2r)}{2(Q^2-r)^2r^{2}},\nonumber\\
	\frac{{d{p_\theta}}}{{d\tau}} &=&-\frac{\partial H_{1}}{\partial\theta}=-\frac{L^{2}\cos\theta}{(Q^{2}r-r^2)\sin^{3}\theta}.\nonumber\\
	{{{\cal H}}_2}:\frac{{dr}}{{d\tau}} &=&\frac{\partial H_{2}}{\partial p_{r}}= {p_r}.\nonumber\\
	{{{\cal H}}_3}:\frac{{dr}}{{d\tau}} &=& \frac{\partial H_{3}}{\partial p_{r}}=-\frac{2}{r}p_{r},
	\frac{{d{p_r}}}{{d\tau}} =-\frac{\partial H_{3}}{\partial r}= -\frac{p_r^{2}}{{r^2}}.\nonumber\\
	{{{\cal H}}_4}:\frac{{d\theta}}{{d\tau}} &=& \frac{\partial H_{4}}{\partial p_{\theta}}=\frac{p_{\theta}}{r(-Q^{2}+r)},
	\frac{{d{p_r}}}{{d\tau}}=-\frac{\partial H_{4}}{\partial r}=\frac{p_\theta^{2}(Q^2-2r)}{{{2r^2(-Q^2+r)^2}}}.\nonumber\\
\end{eqnarray}
Given the initial values of $(r_0, \theta _0, p_{{r_0}}, p_{{\theta _0}})$, the analytic solution for each part can be written as
\begin{eqnarray}
	{{{\cal H}}_1}:{p_r}(\tau) &=& {p_{{r_0}}} + \tau\left( { - \frac{{\partial {{{\cal H}}_1}}}{{\partial r}}} \right),\nonumber\\
	{p_\theta }(\tau) &=& {p_{{\theta _0}}} + \tau\left( { - \frac{{\partial {{{\cal H}}_1}}}{{\partial \theta }}} \right).\nonumber\\
	{{{\cal H}}_2}:r(\tau) &=& {r_0} + \tau{p_{{r_0}}}.\nonumber\\
	{{{\cal H}}_3}:r(\tau) &=& {\left[ {\frac{{{{({r_0}^2 - 3\tau{p_{{r_0}}})}^2}}}{{{r_0}}}} \right]^{\frac{1}{3}}},\nonumber\\
	{p_r}(\tau) &=& {p_{{r_0}}}{\left( {\frac{{{r_0}^2 - 3\tau{p_{{r_0}}}}}{{{r_0}^2}}} \right)^{\frac{1}{3}}}.\nonumber\\
	{{{\cal H}}_4}:\theta (\tau) &=&{\theta _0} + \frac{{\tau{p_{{\theta _0}}}}}{{{r_0}^2-Q^2{r_0}}}\nonumber\\
	{p_r}(\tau) &=& {p_{{r_0}}} - \frac{{\tau({p_{{\theta _0}}}^2Q^2-2{p_{{\theta _0}}}^2{r_0})}}{{{r_0}^3}}.
\end{eqnarray} 
 
Here, $\tau$ denotes the proper time. Defining $h$ as the step size, the $PR{K_6}4$ symplectic algorithm adopts the following difference scheme from step $n-1$ to $n$:
\begin{eqnarray}
	\theta^{*_1} &=&{\theta _{n-1}} + \frac{{c_{12}h{p_{\theta,n-1}}}}{{r_{n-1}^{2}-Q^2{r_{n-1}}}},\nonumber\\
	{p_r}^{*_1}&=& {p_{r,n-1}} - \frac{{c_{12}h(p_{\theta,n-1}^{2}Q^2-2p_{\theta,n-1}^{2}{r_{n-1}})}}{{r_{n-1}^{3}}}.\nonumber\\
	r^{\dagger_1} &=& {\left[ {\frac{{{{(r_{n-1}^{2} - 3c_{12}h{{p_r}^{*_1}})}^2}}}{{{r_{n-1}}}}} \right]^{\frac{1}{3}}},\nonumber\\
	{p_r}^{\dagger_1}&=& {{p_r}^{*_1}}{\left( {\frac{{r_{n-1}^{2} - 3c_{12}h{{p_r}^{*_1}}}}{{r_{n-1}^{2}}}} \right)^{\frac{1}{3}}}.\nonumber\\
	r^{\lozenge_1}&=& r^{\dagger_1} + c_{12}h{p_r}^{\dagger_1}.\nonumber\\
	{p_r}^{\aleph_1}&=& {{p_r}^{\dagger_1}} + c_{12}h\left[-\frac{(-Q^*+Er^{\lozenge_1})(Q^*+Er^{\lozenge_1}-Q^*r^{\lozenge_1})}{{r^{\lozenge_1}}^2(-2+r^{\lozenge_1})^2}+\frac{L^{2}\csc^{2}\theta^{*_1}(-Q^2+2r^{\lozenge_1})}{2(Q^2-r^{\lozenge_1})^2{r^{\lozenge_1}}^{2}}\right],\nonumber\\
	{p_\theta }^{\aleph_1}&=& {p_{\theta,n-1}} + c_{12}h\left[-\frac{L^{2}\cos\theta^{*_1}}{(Q^{2}r^{\lozenge_1}-{r^{\lozenge _1}}^2)\sin^{3}\theta^{*_1}}\right].
\end{eqnarray}
\begin{eqnarray}
	{p_r}^{\aleph_2}&=& {p_r}^{\aleph_1}+ c_{11}h\left[-\frac{(-Q^*+Er^{\lozenge_1})(Q^*+Er^{\lozenge_1}-Q^*r^{\lozenge_1})}{{r^{\lozenge_1}}^2(-2+r^{\lozenge_1})^2}+\frac{L^{2}\csc^{2}\theta^{*_1}(-Q^2+2r^{\lozenge_1})}{2(Q^2-r^{\lozenge_1})^2{r^{\lozenge_1}}^{2}}\right],\nonumber\\
	{p_\theta }^{\aleph_2}&=& {p_\theta }^{\aleph_1} + c_{11}h\left[-\frac{L^{2}\cos\theta^{*_1}}{(Q^{2}r^{\lozenge_1}-{r^{\lozenge _1}}^2)\sin^{3}\theta^{*_1}}\right].\nonumber\\
		r^{\lozenge_2}&=& r^{\lozenge_1}+c_{11}h{p_r}^{\aleph_2}.\nonumber\\
		r^{\dagger_2} &=& {\left[ {\frac{{{{({r^{\lozenge_2}}^{2} - 3c_{11}h{{p_r}^{\aleph_2}})}^2}}}{{{r^{\lozenge_2}}}}} \right]^{\frac{1}{3}}},\nonumber\\
		{p_r}^{\dagger_2}&=& {{p_r}^{\aleph_2}}{\left( {\frac{{{r^{\lozenge_2}}^{2} - 3c_{11}h{{p_r}^{\aleph_2}}}}{{{r^{\lozenge_2}}^{2}}}} \right)^{\frac{1}{3}}}.\nonumber\\
	\theta^{*_2} &=&{\theta^{*_1}} + \frac{{c_{11}h{{p_\theta }^{\aleph_2}}}}{{{r^{\dagger_2}}^{2}-Q^2{r^{\dagger_2}}}},\nonumber\\
	{p_r}^{*_2}&=& {{p_r}^{\dagger_2}} - \frac{{c_{11}h({{p_\theta }^{\aleph_2}}^{2}Q^2-2{{p_\theta }^{\aleph_2}}^{2}{r^{\dagger_2}})}}{{{r^{\dagger_2}}^{3}}}.
\end{eqnarray}
\begin{eqnarray}
	\theta^{*_3} &=&\theta^{*_2}+ \frac{{c_{10}h{{p_\theta}^{\aleph_2}}}}{{{r^{\dagger_2}}^{2}-Q^2{r^{\dagger_2}}}},\nonumber\\
	{p_r}^{*_3}&=& {p_r}^{*_2}- \frac{{c_{10}h({{p_\theta }^{\aleph_2}}^{2}Q^2-2{{p_\theta }^{\aleph_2}}^{2}{r^{\dagger_2}})}}{{{r^{\dagger_2}}^{3}}}.\nonumber\\
	r^{\dagger_3} &=& {\left[ {\frac{{{{({r^{\dagger_2}}^{2} - 3c_{10}h{{p_r}^{*_3}})}^2}}}{{{r^{\dagger_2}}}}} \right]^{\frac{1}{3}}},\nonumber\\
	{p_r}^{\dagger_3}&=& {{p_r}^{*_3}}{\left( {\frac{{{r^{\dagger_2}}^{2} - 3c_{10}h{{p_r}^{*_3}}}}{{{r^{\dagger_2}}^{2}}}} \right)^{\frac{1}{3}}}.\nonumber\\
	r^{\lozenge_3}&=& r^{\dagger_3} + c_{10}h{p_r}^{\dagger_3}.\nonumber\\
	{p_r}^{\aleph_3}&=& {{p_r}^{\dagger_3}} + c_{10}h\left[-\frac{(-Q^*+Er^{\lozenge_3})(Q^*+Er^{\lozenge_3}-Q^*r^{\lozenge_3})}{{r^{\lozenge_3}}^2(-2+r^{\lozenge_3})^2}+\frac{L^{2}\csc^{2}\theta^{*_3}(-Q^2+2r^{\lozenge_3})}{2(Q^2-r^{\lozenge_3})^2{r^{\lozenge_3}}^{2}}\right],\nonumber\\
	{p_\theta }^{\aleph_3}&=&{p_\theta }^{\aleph_2} + c_{10}h\left[-\frac{L^{2}\cos\theta^{*_3}}{(Q^{2}r^{\lozenge_3}-{r^{\lozenge _3}}^2)\sin^{3}\theta^{*_3}}\right].
\end{eqnarray}
\begin{eqnarray}
	{p_r}^{\aleph_4}&=& {p_r}^{\aleph_3}+ c_{9}h\left[-\frac{(-Q^*+Er^{\lozenge_3})(Q^*+Er^{\lozenge_3}-Q^*r^{\lozenge_3})}{{r^{\lozenge_3}}^2(-2+r^{\lozenge_3})^2}+\frac{L^{2}\csc^{2}\theta^{*_3}(-Q^2+2r^{\lozenge_3})}{2(Q^2-r^{\lozenge_3})^2{r^{\lozenge_3}}^{2}}\right],\nonumber\\
	{p_\theta }^{\aleph_4}&=& {p_\theta }^{\aleph_3}+ c_{9}h\left[-\frac{L^{2}\cos\theta^{*_3}}{(Q^{2}r^{\lozenge_3}-{r^{\lozenge _3}}^2)\sin^{3}\theta^{*_3}}\right].\nonumber\\
	r^{\lozenge_4}&=& r^{\lozenge_3}+c_{9}h{p_r}^{\aleph_4}.\nonumber\\
	r^{\dagger_4} &=& {\left[ {\frac{{{{({r^{\lozenge_4}}^{2} - 3c_{9}h{{p_r}^{\aleph_4}})}^2}}}{{{r^{\lozenge_4}}}}} \right]^{\frac{1}{3}}},\nonumber\\
	{p_r}^{\dagger_4}&=& {{p_r}^{\aleph_4}}{\left( {\frac{{{r^{\lozenge_4}}^{2} - 3c_{9}h{{p_r}^{\aleph_4}}}}{{{r^{\lozenge_4}}^{2}}}} \right)^{\frac{1}{3}}}.\nonumber\\
	\theta^{*_4} &=&{\theta^{*_3}} + \frac{{c_{9}h{{p_\theta }^{\aleph_4}}}}{{{r^{\dagger_4}}^{2}-Q^2{r^{\dagger_4}}}},\nonumber\\
	{p_r}^{*_4}&=& {{p_r}^{\dagger_4}} - \frac{{c_{9}h({{p_\theta }^{\aleph_4}}^{2}Q^2-2{{p_\theta }^{\aleph_4}}^{2}{r^{\dagger_4}})}}{{{r^{\dagger_4}}^{3}}}.
\end{eqnarray}
\begin{eqnarray}
	\theta^{*_5} &=&\theta^{*_4} + \frac{{c_{8}h{{p_\theta }^{\aleph_4}}}}{{{r^{\dagger_4}}^{2}-Q^2{r^{\dagger_4}}}},\nonumber\\
	{p_r}^{*_5}&=& {p_r}^{*_4}- \frac{{c_{8}h({{p_\theta }^{\aleph_4}}^{2}Q^2-2{{p_\theta }^{\aleph_4}}^{2}{r^{\dagger_4}})}}{{{r^{\dagger_4}}^{3}}}.\nonumber\\
	r^{\dagger_5} &=& {\left[ {\frac{{{{({r^{\dagger_4}}^{2} - 3c_{8}h{{p_r}^{*_5}})}^2}}}{{{r^{\dagger_4}}}}} \right]^{\frac{1}{3}}},\nonumber\\
	{p_r}^{\dagger_5}&=& {{p_r}^{*_5}}{\left( {\frac{{{r^{\dagger_4}}^{2} - 3c_{8}h{{p_r}^{*_5}}}}{{{r^{\dagger_4}}^{2}}}} \right)^{\frac{1}{3}}}.\nonumber\\
	r^{\lozenge_5}&=& r^{\dagger_5} + c_{8}h{p_r}^{\dagger_5}.\nonumber\\
	{p_r}^{\aleph_5}&=& {{p_r}^{\dagger_5}} + c_{8}h\left[-\frac{(-Q^*+Er^{\lozenge_5})(Q^*+Er^{\lozenge_5}-Q^*r^{\lozenge_5})}{{r^{\lozenge_5}}^2(-2+r^{\lozenge_5})^2}+\frac{L^{2}\csc^{2}\theta^{*_5}(-Q^2+2r^{\lozenge_5})}{2(Q^2-r^{\lozenge_5})^2{r^{\lozenge_5}}^{2}}\right],\nonumber\\
	{p_\theta }^{\aleph_5}&=&{p_\theta }^{\aleph_4} + c_{8}h\left[-\frac{L^{2}\cos\theta^{*_5}}{(Q^{2}r^{\lozenge_5}-{r^{\lozenge _5}}^2)\sin^{3}\theta^{*_5}}\right].
\end{eqnarray}
\begin{eqnarray}
	{p_r}^{\aleph_6}&=& {p_r}^{\aleph_5}+ c_{7}h\left[-\frac{(-Q^*+Er^{\lozenge_5})(Q^*+Er^{\lozenge_5}-Q^*r^{\lozenge_5})}{{r^{\lozenge_5}}^2(-2+r^{\lozenge_5})^2}+\frac{L^{2}\csc^{2}\theta^{*_5}(-Q^2+2r^{\lozenge_5})}{2(Q^2-r^{\lozenge_5})^2{r^{\lozenge_5}}^{2}}\right],\nonumber\\
	{p_\theta }^{\aleph_6}&=& {p_\theta }^{\aleph_5}+ c_{7}h\left[-\frac{L^{2}\cos\theta^{*_5}}{(Q^{2}r^{\lozenge_5}-{r^{\lozenge _5}}^2)\sin^{3}\theta^{*_5}}\right].\nonumber\\
	r^{\lozenge_6}&=& r^{\lozenge_5}+c_{7}h{p_r}^{\aleph_6}.\nonumber\\
	r^{\dagger_6} &=& {\left[ {\frac{{{{({r^{\lozenge_6}}^{2} - 3c_{7}h{{p_r}^{\aleph_6}})}^2}}}{{{r^{\lozenge_6}}}}} \right]^{\frac{1}{3}}},\nonumber\\
	{p_r}^{\dagger_6}&=& {{p_r}^{\aleph_6}}{\left( {\frac{{{r^{\lozenge_6}}^{2} - 3c_{7}h{{p_r}^{\aleph_6}}}}{{{r^{\lozenge_6}}^{2}}}} \right)^{\frac{1}{3}}}.\nonumber\\
	\theta^{*_6} &=&{\theta^{*_5}} + \frac{{c_{7}h{{p_\theta }^{\aleph_6}}}}{{{r^{\dagger_6}}^{2}-Q^2{r^{\dagger_6}}}},\nonumber\\
	{p_r}^{*_6}&=& {{p_r}^{\dagger_6}} - \frac{{c_{7}h({{p_\theta }^{\aleph_6}}^{2}Q^2-2{{p_\theta }^{\aleph_6}}^{2}{r^{\dagger_6}})}}{{{r^{\dagger_6}}^{3}}}.
\end{eqnarray}
\begin{eqnarray}
	\theta^{*_7} &=&\theta^{*_6} + \frac{{c_{6}h{{p_\theta }^{\aleph_6}}}}{{{r^{\dagger_6}}^{2}-Q^2{r^{\dagger_6}}}},\nonumber\\
	{p_r}^{*_7}&=& {p_r}^{*_6}- \frac{{c_{6}h({{p_\theta }^{\aleph_6}}^{2}Q^2-2{{p_\theta }^{\aleph_6}}^{2}{r^{\dagger_6}})}}{{{r^{\dagger_6}}^{3}}}.\nonumber\\
	r^{\dagger_7} &=& {\left[ {\frac{{{{({r^{\dagger_6}}^{2} - 3c_{6}h{{p_r}^{*_7}})}^2}}}{{{r^{\dagger_6}}}}} \right]^{\frac{1}{3}}},\nonumber\\
	{p_r}^{\dagger_7}&=& {{p_r}^{*_7}}{\left( {\frac{{{r^{\dagger_6}}^{2} - 3c_{6}h{{p_r}^{*_7}}}}{{{r^{\dagger_6}}^{2}}}} \right)^{\frac{1}{3}}}.\nonumber\\
	r^{\lozenge_7}&=& r^{\dagger_7} + c_{6}h{p_r}^{\dagger_7}.\nonumber\\
	{p_r}^{\aleph_7}&=& {{p_r}^{\dagger_7}} + c_{6}h\left[-\frac{(-Q^*+Er^{\lozenge_7})(Q^*+Er^{\lozenge_7}-Q^*r^{\lozenge_7})}{{r^{\lozenge_7}}^2(-2+r^{\lozenge_7})^2}+\frac{L^{2}\csc^{2}\theta^{*_7}(-Q^2+2r^{\lozenge_7})}{2(Q^2-r^{\lozenge_7})^2{r^{\lozenge_7}}^{2}}\right],\nonumber\\
	{p_\theta }^{\aleph_7}&=&{p_\theta }^{\aleph_6} + c_{6}h\left[-\frac{L^{2}\cos\theta^{*_7}}{(Q^{2}r^{\lozenge_7}-{r^{\lozenge _7}}^2)\sin^{3}\theta^{*_7}}\right].
\end{eqnarray}
\begin{eqnarray}
	{p_r}^{\aleph_8}&=& {p_r}^{\aleph_7}+ c_{5}h\left[-\frac{(-Q^*+Er^{\lozenge_7})(Q^*+Er^{\lozenge_7}-Q^*r^{\lozenge_7})}{{r^{\lozenge_7}}^2(-2+r^{\lozenge_7})^2}+\frac{L^{2}\csc^{2}\theta^{*_7}(-Q^2+2r^{\lozenge_7})}{2(Q^2-r^{\lozenge_7})^2{r^{\lozenge_7}}^{2}}\right],\nonumber\\
	{p_\theta }^{\aleph_8}&=&{p_\theta }^{\aleph_7}+ c_{5}h\left[-\frac{L^{2}\cos\theta^{*_7}}{(Q^{2}r^{\lozenge_7}-{r^{\lozenge _7}}^2)\sin^{3}\theta^{*_7}}\right].\nonumber\\
	r^{\lozenge_8}&=& r^{\lozenge_7}+c_{5}h{p_r}^{\aleph_8}.\nonumber\\
	r^{\dagger_8} &=& {\left[ {\frac{{{{({r^{\lozenge_8}}^{2} - 3c_{5}h{{p_r}^{\aleph_8}})}^2}}}{{{r^{\lozenge_8}}}}} \right]^{\frac{1}{3}}},\nonumber\\
	{p_r}^{\dagger_8}&=& {{p_r}^{\aleph_8}}{\left( {\frac{{{r^{\lozenge_8}}^{2} - 3c_{5}h{{p_r}^{\aleph_8}}}}{{{r^{\lozenge_8}}^{2}}}} \right)^{\frac{1}{3}}}.\nonumber\\
	\theta^{*_8}&=&{\theta^{*_7}} + \frac{{c_{5}h{{p_\theta }^{\aleph_8}}}}{{{r^{\dagger_8}}^{2}-Q^2{r^{\dagger_8}}}},\nonumber\\
	{p_r}^{*_8}&=& {{p_r}^{\dagger_8}} - \frac{{c_{5}h({{p_\theta }^{\aleph_8}}^{2}Q^2-2{{p_\theta }^{\aleph_8}}^{2}{r^{\dagger_8}})}}{{{r^{\dagger_8}}^{3}}}.
\end{eqnarray}
\begin{eqnarray}
	\theta^{*_9} &=&\theta^{*_8}+ \frac{{c_{4}h{{p_\theta }^{\aleph_8}}}}{{{r^{\dagger_8}}^{2}-Q^2{r^{\dagger_8}}}},\nonumber\\
	{p_r}^{*_9}&=& {p_r}^{*_8}- \frac{{c_{4}h({{p_\theta }^{\aleph_8}}^{2}Q^2-2{{p_\theta }^{\aleph_8}}^{2}{r^{\dagger_8}})}}{{{r^{\dagger_8}}^{3}}}.\nonumber\\
	r^{\dagger_9} &=& {\left[ {\frac{{{{({r^{\dagger_8}}^{2} - 3c_{4}h{{p_r}^{*_9}})}^2}}}{{{r^{\dagger_8}}}}} \right]^{\frac{1}{3}}},\nonumber\\
	{p_r}^{\dagger_9}&=& {{p_r}^{*_9}}{\left( {\frac{{{r^{\dagger_8}}^{2} - 3c_{4}h{{p_r}^{*_9}}}}{{{r^{\dagger_8}}^{2}}}} \right)^{\frac{1}{3}}}.\nonumber\\
	r^{\lozenge_9}&=& r^{\dagger_9} + c_{4}h{p_r}^{\dagger_9}.\nonumber\\
	{p_r}^{\aleph_9}&=& {{p_r}^{\dagger_9}} + c_{4}h\left[-\frac{(-Q^*+Er^{\lozenge_9})(Q^*+Er^{\lozenge_9}-Q^*r^{\lozenge_9})}{{r^{\lozenge_9}}^2(-2+r^{\lozenge_9})^2}+\frac{L^{2}\csc^{2}\theta^{*_9}(-Q^2+2r^{\lozenge_9})}{2(Q^2-r^{\lozenge_9})^2{r^{\lozenge_9}}^{2}}\right],\nonumber\\
	{p_\theta }^{\aleph_9}&=&{p_\theta }^{\aleph_8} + c_{4}h\left[-\frac{L^{2}\cos\theta^{*_9}}{(Q^{2}r^{\lozenge_9}-{r^{\lozenge _9}}^2)\sin^{3}\theta^{*_9}}\right].
\end{eqnarray}
\begin{eqnarray}
	{p_r}^{\aleph_{10}}&=& {p_r}^{\aleph_9}+ c_{3}h\left[-\frac{(-Q^*+Er^{\lozenge_9})(Q^*+Er^{\lozenge_9}-Q^*r^{\lozenge_9})}{{r^{\lozenge_9}}^2(-2+r^{\lozenge_9})^2}+\frac{L^{2}\csc^{2}\theta^{*_9}(-Q^2+2r^{\lozenge_9})}{2(Q^2-r^{\lozenge_9})^2{r^{\lozenge_9}}^{2}}\right],\nonumber\\
	{p_\theta }^{\aleph_{10}}&=&{p_\theta }^{\aleph_9}+ c_{3}h\left[-\frac{L^{2}\cos\theta^{*_9}}{(Q^{2}r^{\lozenge_9}-{r^{\lozenge _9}}^2)\sin^{3}\theta^{*_9}}\right].\nonumber\\
	r^{\lozenge_{10}}&=& r^{\lozenge_9}+c_{3}h{p_r}^{\aleph_{10}}.\nonumber\\
	r^{\dagger_{10}} &=& {\left[ {\frac{{{{({r^{\lozenge_{10}}}^{2} - 3c_{3}h{{p_r}^{\aleph_{10}}})}^2}}}{{{r^{\lozenge_{10}}}}}} \right]^{\frac{1}{3}}},\nonumber\\
	{p_r}^{\dagger_{10}}&=& {{p_r}^{\aleph_{10}}}{\left( {\frac{{{r^{\lozenge_{10}}}^{2} - 3c_{3}h{{p_r}^{\aleph_{10}}}}}{{{r^{\lozenge_{10}}}^{2}}}} \right)^{\frac{1}{3}}}.\nonumber\\
	\theta^{*_{10}}&=&{\theta^{*_9}} + \frac{{c_{3}h{{p_\theta }^{\aleph_{10}}}}}{{{r^{\dagger_{10}}}^{2}-Q^2{r^{\dagger_{10}}}}},\nonumber\\
	{p_r}^{*_{10}}&=& {{p_r}^{\dagger_{10}}} - \frac{{c_{3}h({{p_\theta }^{\aleph_{10}}}^{2}Q^2-2{{p_\theta }^{\aleph_{10}}}^{2}{r^{\dagger_{10}}})}}{{{r^{\dagger_{10}}}^{3}}}.
\end{eqnarray}
\begin{eqnarray}
	\theta^{*_{11}} &=&\theta^{*_{10}}+ \frac{{c_{2}h{{p_\theta }^{\aleph_{10}}}}}{{{r^{\dagger_{10}}}^{2}-Q^2{r^{\dagger_{10}}}}},\nonumber\\
	{p_r}^{*_{11}}&=& {p_r}^{*_{10}}- \frac{{c_{2}h({{p_\theta }^{\aleph_{10}}}^{2}Q^2-2{{p_\theta }^{\aleph_{10}}}^{2}{r^{\dagger_{10}}})}}{{{r^{\dagger_{10}}}^{3}}}.\nonumber\\
	r^{\dagger_{11}} &=& {\left[ {\frac{{{{({r^{\dagger_{10}}}^{2} - 3c_{2}h{{p_r}^{*_{11}}})}^2}}}{{{r^{\dagger_{10}}}}}} \right]^{\frac{1}{3}}},\nonumber\\
	{p_r}^{\dagger_{11}}&=& {{p_r}^{*_{11}}}{\left( {\frac{{{r^{\dagger_{10}}}^{2} - 3c_{2}h{{p_r}^{*_{11}}}}}{{{r^{\dagger_{10}}}^{2}}}} \right)^{\frac{1}{3}}}.\nonumber\\
	r^{\lozenge_{11}}&=& r^{\dagger_{11}} + c_{2}h{p_r}^{\dagger_{11}}.\nonumber\\
	{p_r}^{\aleph_{11}}&=& {{p_r}^{\dagger_{11}}} + c_{2}h\left[-\frac{(-Q^*+Er^{\lozenge_{11}})(Q^*+Er^{\lozenge_{11}}-Q^*r^{\lozenge_{11}})}{{r^{\lozenge_{11}}}^2(-2+r^{\lozenge_{11}})^2}+\frac{L^{2}\csc^{2}\theta^{*_{11}}(-Q^2+2r^{\lozenge_{11}})}{2(Q^2-r^{\lozenge_{11}})^2{r^{\lozenge_{11}}}^{2}}\right],\nonumber\\
	{p_\theta }^{\aleph_{11}}&=&{p_\theta }^{\aleph_{10}} + c_{2}h\left[-\frac{L^{2}\cos\theta^{*_{11}}}{(Q^{2}r^{\lozenge_{11}}-{r^{\lozenge _{11}}}^2)\sin^{3}\theta^{*_{11}}}\right].
\end{eqnarray}
\begin{eqnarray}
	{p_r}^{\aleph_{12}}&=& {p_r}^{\aleph_{11}}+ c_{1}h\left[-\frac{(-Q^*+Er^{\lozenge_{11}})(Q^*+Er^{\lozenge_{11}}-Q^*r^{\lozenge_{11}})}{{r^{\lozenge_{11}}}^2(-2+r^{\lozenge_{11}})^2}+\frac{L^{2}\csc^{2}\theta^{*_{11}}(-Q^2+2r^{\lozenge_{11}})}{2(Q^2-r^{\lozenge_{11}})^2{r^{\lozenge_{11}}}^{2}}\right],\nonumber\\
	{p_\theta }^{\aleph_{12}}&=&{p_\theta }^{\aleph_{11}}+ c_{1}h\left[-\frac{L^{2}\cos\theta^{*_{11}}}{(Q^{2}r^{\lozenge_{11}}-{r^{\lozenge _{11}}}^2)\sin^{3}\theta^{*_{11}}}\right].\nonumber\\
	r^{\lozenge_{12}}&=& r^{\lozenge_{11}}+c_{1}h{p_r}^{\aleph_{12}}.\nonumber\\
	r^{\dagger_{12}} &=& {\left[ {\frac{{{{({r^{\lozenge_{12}}}^{2} - 3c_{1}h{{p_r}^{\aleph_{12}}})}^2}}}{{{r^{\lozenge_{12}}}}}} \right]^{\frac{1}{3}}},\nonumber\\
	{p_r}^{\dagger_{12}}&=& {{p_r}^{\aleph_{12}}}{\left( {\frac{{{r^{\lozenge_{12}}}^{2} - 3c_{1}h{{p_r}^{\aleph_{12}}}}}{{{r^{\lozenge_{12}}}^{2}}}} \right)^{\frac{1}{3}}}.\nonumber\\
	\theta^{*_{12}}&=&{\theta^{*_{11}}} + \frac{{c_{1}h{{p_\theta }^{\aleph_{12}}}}}{{{r^{\dagger_{12}}}^{2}-Q^2{r^{\dagger_{12}}}}},\nonumber\\
	{p_r}^{*_{12}}&=& {{p_r}^{\dagger_{12}}} - \frac{{c_{1}h({{p_\theta }^{\aleph_{12}}}^{2}Q^2-2{{p_\theta }^{\aleph_{12}}}^{2}{r^{\dagger_{12}}})}}{{{r^{\dagger_{12}}}^{3}}}.
\end{eqnarray}
\begin{eqnarray}
	r_n&=&r^{\dagger_{12}},\nonumber\\
	\theta_n&=&\theta^{*_{12}},\nonumber\\
	{p_{r,n}}&=& {p_r}^{*_{12}},\nonumber\\
	{p_{\theta,n}}&=&{p_\theta }^{\aleph_{12}}.
\end{eqnarray}



\begin{thebibliography}{99}
\bibitem{Banerjee:2020ubc}
I.~Banerjee, B.~Mandal and S.~SenGupta,
``Signatures of Einstein-Maxwell dilaton-axion gravity from the observed jet power and the radiative efficiency,''
\textit{Phys. Rev. D} \textbf{103}, no.4, 044046 (2021)
doi:10.1103/PhysRevD.103.044046
[arXiv:2007.03947 [gr-qc]].

\bibitem{Feng:2023iha}
H.~Feng, Y.~Wu, R.~J.~Yang and L.~Modesto,
``Choked accretion onto Kerr-Sen black holes in Einstein-Maxwell-dilaton-axion gravity,''
\textit{Phys. Rev. D} \textbf{109}, no.6, 063014 (2024)
doi:10.1103/PhysRevD.109.063014
[arXiv:2301.02779 [astro-ph.HE]].

\bibitem{Feng:2024iqj}
H.~Feng, R.~J.~Yang and W.~Q.~Chen,
``Thin accretion disk and shadow of Kerr{\textendash}Sen black hole in Einstein{\textendash}Maxwell-dilaton{\textendash}axion gravity,''
\textit{Astropart. Phys.} \textbf{166}, 103075 (2025)
doi:10.1016/j.astropartphys.2024.103075
[arXiv:2403.18541 [gr-qc]].

\bibitem{An:2017hby}
J.~An, J.~Peng, Y.~Liu and X.~H.~Feng,
``Kerr-Sen Black Hole as Accelerator for Spinning Particles,''
\textit{Phys. Rev. D} \textbf{97}, no.2, 024003 (2018)
doi:10.1103/PhysRevD.97.024003
[arXiv:1710.08630 [gr-qc]].

\bibitem{Wu:2003}
S.~Q.~Wu and X.~Cai,
``Massive complex scalar field in the Kerr–Sen geometry: Exact solution of wave equation and Hawking radiation,''
\textit{J. Math. Phys.} \textbf{44}, no.3, 1084-1088 (2003)
doi:10.1063/1.1539899
[arXiv:gr-qc/0303075 [gr-qc]].

\bibitem{Vieira:2018hij}
H.~S.~Vieira and V.~B.~Bezerra,
``Charged scalar fields in a Kerr\textendash{}Sen black hole: exact solutions, Hawking radiation, and resonant frequencies,''
\textit{Chin. Phys. C} \textbf{43}, no.3, 035102 (2019)
doi:10.1088/1674-1137/43/3/035102
[arXiv:1811.06129 [gr-qc]].

\bibitem{Wu:2021pgf}
X.~Wu and X.~Zhang,
``Connections between the Shadow Radius and the Quasinormal Modes of Kerr-Sen Black Hole,''
\textit{Universe} \textbf{8}, no.11, 604 (2022)
doi:10.3390/universe8110604
[arXiv:2112.11066 [gr-qc]].

\bibitem{Cardoso:2008bp}
V.~Cardoso, A.~S.~Miranda, E.~Berti, H.~Witek and V.~T.~Zanchin,
``Geodesic stability, Lyapunov exponents and quasinormal modes,''
\textit{Phys. Rev. D} \textbf{79}, no.6, 064016 (2009)
doi:10.1103/PhysRevD.79.064016
[arXiv:0812.1806 [hep-th]].

\bibitem{Narang:2020bgo}
A.~Narang, S.~Mohanty and A.~Kumar,
``Test of Kerr-Sen metric with black hole observations,''
[arXiv:2002.12786 [gr-qc]].

\bibitem{GM} 
G.~W.~Gibbons and K.~I.~Maeda,
``Black Holes and Membranes in Higher Dimensional Theories with Dilaton Fields,''
\textit{Nucl. Phys. B} \textbf{298}, 741 (1988)
doi:10.1016/0550-3213(88)90006-5

\bibitem{Garfinkle:1990qj}
D.~Garfinkle, G.~T.~Horowitz and A.~Strominger,
``Charged black holes in string theory,''
\textit{Phys. Rev. D} \textbf{43}, 3140 (1991)
[erratum: \textit{Phys. Rev. D} \textbf{45}, 3888 (1992)]
doi:10.1103/PhysRevD.43.3140

\bibitem{Sen:1992ua}
A.~Sen,
``Rotating charged black hole solution in heterotic string theory,''
\textit{Phys. Rev. Lett.} \textbf{69}, 1006-1009 (1992)
doi:10.1103/PhysRevLett.69.1006
[arXiv:hep-th/9204046 [hep-th]].

\bibitem{Soroushfar:2016yea}
S.~Soroushfar, R.~Saffari and E.~Sahami,
``Geodesic equations in the static and rotating dilaton black holes: Analytical solutions and applications,''
\textit{Phys. Rev. D} \textbf{94}, no.2, 024010 (2016)
doi:10.1103/PhysRevD.94.024010
[arXiv:1601.03143 [gr-qc]].

\bibitem{Blaga:2014spa}
C.~Blaga,
``Timelike geodesics around a charged spherically symmetric dilaton black hole,''
\textit{Serb. Astron. J.} \textbf{190}, 41 (2015)
doi:10.2298/SAJ1590041B
[arXiv:1407.1504 [gr-qc]].

\bibitem{Fernando:2011ki}
S.~Fernando,
``Null Geodesics of Charged Black Holes in String Theory,''
\textit{Phys. Rev. D} \textbf{85}, 024033 (2012)
doi:10.1103/PhysRevD.85.024033
[arXiv:1109.0254 [hep-th]].

\bibitem{Lungu:2024ewz}
V.~Lungu, M.~A.~Dariescu, C.~Dariescu and C.~Stelea,
``Charged Particles Moving around a Spherically Symmetric Dilatonic Black Hole,''
\textit{Adv. High Energy Phys.} \textbf{2024}, 6666609 (2024)
doi:10.1155/2024/6666609
[arXiv:2401.17398 [hep-th]].

\bibitem{MA2022epjc}
D.~Z.~Ma, F.~Xia, D.~Zhang, G.~Y.~Fu and J.~P.~Wu,
``Chaotic dynamics of string around the conformal black hole,''
\textit{Eur. Phys. J. C} \textbf{82}, 372 (2022)
doi:10.1140/epjc/s10052-022-10338-5.

\bibitem{MA2020JHEP}
D.~Z.~Ma, D.~Zhang, G.~Y.~Fu and J.~P.~Wu,
``Chaotic dynamics of string around charged black brane with hyperscaling violation,''
\textit{JHEP} \textbf{01}, 103 (2020)
doi:10.1007/JHEP01(2020)103
[arXiv:1911.09913 [hep-th]].

\bibitem{Mss2016}
J.~Maldacena, D.~Stanford, 
``A bound on chaos,''
\textit{JHEP} \textbf{106}, 1608(2022)
doi:10.1007/JHEP08/282016/29106
[arXiv:1503.01409 [hep-th]].

\bibitem{MA2014PRD}
D.~Z.~Ma, J.~P.~Wu, J.~F.~Zhang,
``Chaos from the ring string in Gauss-Bonnet black hole in $ADS5$ space,''
\textit{Phys. Rev. D} \textbf{89}, 086011 (2014)
doi:10.1103/PhysRevD.89.086011.
[arXiv:1405.3563 [hep-th]] .

\bibitem{MA2008216}
D.~Z.~Ma, X.~Wu and J.~F.~Zhu,
``Velocity scaling method to correct individual Kepler energies,''
\textit{New Astron.} \textbf{13}, no. 4, 216-223 (2008)
doi:10.1016/j.newast.2007.09.002.

\bibitem{Wang:2021gja}
Y.~Wang, W.~Sun, F.~Y.~Liu and X.~Wu,
``Construction of Explicit Symplectic Integrators in General Relativity. I. Schwarzschild Black Holes,''
\textit{Astrophys. J.} \textbf{907}, no. 2, 66 (2021)
doi:10.3847/1538-4357/abcb8d
[arXiv:2102.00373 [gr-qc]].

\bibitem{Wang:2021xww}
Y.~Wang, W.~Sun, F.~Y.~Liu and X.~Wu,
``Construction of Explicit Symplectic Integrators in General Relativity. II. Reissner--Nordström Black Holes,''
\textit{Astrophys. J.} \textbf{909}, no. 1, 22 (2021)
doi:10.3847/1538-4357/abd701
[arXiv:2103.02864 [gr-qc]].

\bibitem{Wang:2021yqk}
Y.~Wang, W.~Sun, F.~Y.~Liu and X.~Wu,
``Construction of Explicit Symplectic Integrators in General Relativity. III. Reissner--Nordström-(anti)-de Sitter Black Holes,''
\textit{Astrophys. J. Suppl.} \textbf{254}, no. 1, 8 (2021)
doi:10.3847/1538-4365/abf116
[arXiv:2103.12272 [gr-qc]].

\bibitem{Gerald:1988}
G.~J.~Sussman and J.~Wisdom,
``Numerical Evidence That the Motion of Pluto Is Chaotic,''
\textit{Science} \textbf{241}, 433-437 (1988)
doi:10.1126/science.241.4864.433

\bibitem{Li:2018wtz}
D.~Li and X.~Wu,
``Chaotic motion of neutral and charged particles in a magnetized Ernst-Schwarzschild spacetime,''
\textit{Eur. Phys. J. Plus} \textbf{134}, no.3, 96 (2019)
doi:10.1140/epjp/i2019-12502-9
[arXiv:1803.02119 [gr-qc]].

\bibitem{Zhang:2023lrt}
L.~Zhang, S.~Chen, Q.~Pan and J.~Jing,
``Chaotic motion of scalar particle coupling to Chern\textendash{}Simons invariant in the stationary axisymmetric Einstein\textendash{}Maxwell dilaton black hole spacetime,''
\textit{Eur. Phys. J. C} \textbf{83}, no.9, 828 (2023)
doi:10.1140/epjc/s10052-023-12008-6
[arXiv:2309.12604 [gr-qc]].

\bibitem{Cao:2024bjk}
W.~F.~Cao, Y.~Huang and H.~S.~Zhang,
``Screen chaotic motion by Shannon entropy in curved spacetimes,''
\textit{Eur. Phys. J. C} \textbf{85}, no.5, 568 (2025) 
doi:10.1140/epjc/s10052-025-14310-x
[arXiv:2410.20870 [gr-qc]].

\bibitem{Xu:2024ble}
Z.~M.~Xu, D.~Z.~Ma, W.~F.~Cao and K.~Li,
``Chaotic motion of the charged test particle in Kerr-MOG black hole with explicit symplectic algorithms,''
\textit{Eur. Phys. J. C} \textbf{85}, no.7, 770 (2025)
doi:10.1140/epjc/s10052-025-14425-1 
[arXiv:2412.06122 [gr-qc]].

\bibitem{Zhou_2022}
N.~Y.~Zhou, H.~X.~Zhang, W.~F.~Liu and X.~Wu,
``A Note on the Construction of Explicit Symplectic Integrators for Schwarzschild Spacetimes,''
\textit{Astrophys. J.} \textbf{927}, no. 2, 160 (2022)
doi:10.3847/1538-4357/ac497f.

\bibitem{Wu:2003pe}
X.~Wu and T.~Y.~Huang,
``Computation of Lyapunov exponents in general relativity,''
\textit{Phys. Lett. A} \textbf{313}, 77-81 (2003)
doi:10.1016/S0375-9601(03)00720-5
[arXiv:gr-qc/0302118 [gr-qc]].

\bibitem{2000CeMDA..78..167F}
C.~Froeschlé and E.~Lega,
``On the Structure of Symplectic Mappings. The Fast Lyapunov Indicator: a Very Sensitive Tool,''
\textit{Celest. Mech. Dyn. Astron.} \textbf{78}, 167-195 (2000)
doi:10.1023/A:1011141018230.

\bibitem{Wu:2006rx}
X.~Wu, T.~Y.~Huang and H.~Zhang,
``Lyapunov indices with two nearby trajectories in a curved spacetime,''
\textit{Phys. Rev. D} \textbf{74}, 083001 (2006)
doi:10.1103/PhysRevD.74.083001
[arXiv:1006.5251 [gr-qc]].


		
\end{thebibliography}
\end{document}